\documentclass[12pt]{article}

\usepackage{amsthm,amsmath,amsfonts,amssymb}
\usepackage{graphicx,psfrag,epsf}
\usepackage{enumerate}
\usepackage{natbib}
\usepackage{mathtools}
\usepackage{booktabs}
\usepackage{color}
\usepackage[english]{babel} % English language/hyphenation
\usepackage[protrusion=true,expansion=true]{microtype} % Better typography
\usepackage{amsmath,amsfonts,amsthm}
\usepackage{dsfont}
\usepackage{amssymb}
\usepackage{chngcntr}
\usepackage[toc,page]{appendix}
\usepackage{textcomp}
\usepackage{url}
\usepackage[colorlinks,citecolor=blue,urlcolor=blue,backref=page,backref=page]{hyperref}

\usepackage{enumitem}
\newlist{steps}{enumerate}{1}
\setlist[steps, 1]{label = Step \arabic*:}
\usepackage{array}
\newcolumntype{P}[1]{>{\centering\arraybackslash}p{#1}}
\usepackage{xurl}
\newtheorem{prop}{Proposition}
\usepackage{xr}
\usepackage{graphicx}
\usepackage{natbib} %comment out if you do not have the package
\usepackage{url} % not crucial - just used below for the URL 
\usepackage[ruled,vlined]{algorithm2e}
\usepackage{algorithmic}

\usepackage{array}
\usepackage{booktabs} % Horizontal rules in tables
\usepackage{setspace}
\usepackage{amsmath}
\usepackage{xcolor}
\usepackage{soul}
\usepackage{multirow}
\usepackage{enumitem}
\usepackage{microtype} % Slightly tweak font spacing for aesthetics
\usepackage[font = small,labelfont=bf,textfont=it]{subcaption}
\usepackage{xcolor}
\usepackage{tabularx}
\usepackage[font = small,labelfont=bf,textfont=it]{caption} % Custom captions under/above floats in tables or figures
\usepackage{footnote}
\usepackage{etoolbox}
\usepackage{tabularx}
\usepackage{authblk}
\usepackage{caption}
\usepackage{indentfirst}
\usepackage{enumitem}
\usepackage{textcomp}
\usepackage{graphicx}
\usepackage{caption}
\usepackage{subcaption}
\DeclareUnicodeCharacter{2113}{}
\DeclareUnicodeCharacter{2082}{}

\makeatletter
\def\BState{\State\hskip-\ALG@thistlm}

\newcommand{\distas}[1]{\mathbin{\overset{#1}{\kern\z@\sim}}}%
\newcommand{\bm}[1]{\mathbf{#1}}

\newsavebox{\mybox}\newsavebox{\mysim}
\newcommand{\distras}[1]{%
  \savebox{\mybox}{\hbox{\kern3pt$\scriptstyle#1$\kern3pt}}%
  \savebox{\mysim}{\hbox{$\sim$}}%
  \mathbin{\overset{#1}{\kern\z@\resizebox{\wd\mybox}{\ht\mysim}{$\sim$}}}%
}

\newcommand{\be}{\begin{equation}}
\newcommand{\ee}{\end{equation}}
    \newcommand{\bi}{\begin{itemize}}
\newcommand{\ei}{\end{itemize}}
\newcommand{\ben}{\begin{enumerate}}
\newcommand{\een}{\end{enumerate}}

\newcolumntype{K}[1]{\geq {\centering\arraybackslash}p{#1}}
\allowdisplaybreaks

\makeatother

\let\oldbibliography\thebibliography
\renewcommand{\thebibliography}[1]{\oldbibliography{#1}
\setlength{\itemsep}{0pt}} %Reducing spacing in the bibliography.

\newcommand{\blind}{1}

\patchcmd{\footnotemark}{\stepcounter{footnote}}{\refstepcounter{footnote}}{}{}
\setcitestyle{square}

\usepackage{tikz}
\newcommand*\circled[1]{\tikz[baseline=(char.base)]{
            \node[shape=circle,draw,inner sep=0.6pt] (char) {#1};}}

\newcommand{\ilink}{\hyperlink{ilink}{(i)}\xspace}
\newcommand{\iilink}{\hyperlink{iilink}{(ii)}\xspace}
\newcommand{\iiilink}{\hyperlink{iiilink}{(iii)}\xspace}

\begin{document}

\def\spacingset#1{\renewcommand{\baselinestretch}%
{#1}\small\normalsize} \spacingset{1}

\if1\blind
{
  \title{\bf BLAST: Bayesian online change-point detection with structured image data}
  \small
   \author{Xiaojun Zheng\footnote{Department of Statistical Science, Duke University}\;, Simon Mak$^*$\footnote{The authors were supported by NSF Grants CSSI 2004571, DMS 2220496, DMS 2316012, and Department of Energy Grant DE-SC0024477.}
   }
  \maketitle
} \fi

\if0\blind
{
  \bigskip
  \bigskip
  \bigskip
  \begin{center}
    {\LARGE\bf Taguchi meets Gaussian Process: Transformed Analysis of Marginal Mean in Continuous Space}
\end{center}

  \medskip
} \fi

\begin{abstract}
The prompt online detection of abrupt changes in image data is essential for timely decision-making in broad applications, from video surveillance to manufacturing quality control. Existing methods, however, face three key challenges. First, the high-dimensional nature of image data introduces computational bottlenecks for efficient real-time monitoring. Second, changes often involve structural image features, e.g., edges, blurs and/or shapes, and ignoring such structure can lead to delayed change detection. Third, existing methods are largely non-Bayesian and thus do not provide a quantification of monitoring uncertainty for confident detection. We address this via a novel Bayesian onLine Structure-Aware change deTection (BLAST) method. BLAST first leverages a deep Gaussian Markov random field prior to elicit desirable image structure from offline reference data. With this prior elicited, BLAST employs a new Bayesian online change-point procedure for image monitoring via its so-called posterior run length distribution. This posterior run length distribution can be computed in an online fashion using $\mathcal{O}(p^2)$ work at each time-step, where $p$ is the number of image pixels; this facilitates scalable Bayesian online monitoring of large images. We demonstrate the effectiveness of BLAST over existing methods in a suite of numerical experiments and in two applications, the first on street scene monitoring and the second on real-time process monitoring for metal additive manufacturing.
\end{abstract}

\vfill

\newpage
\spacingset{1.55} % DON'T change the spacing!

\section{Introduction}
\label{sec:intro}

The need for prompt online detection of abrupt changes in image data arises in a myriad of modern applications, from video surveillance \citep{He2015DeepRL} to manufacturing quality control \citep{zhang2022review}. For such applications, there are three critical challenges that can hinder scalable and quick change-point detection for timely decision-making. Consider, e.g., the real-time monitoring of a street block, where one wishes to quickly detect any sudden disturbances. The first Challenge \hypertarget{ilink}{}\ilink arises from the \textit{high-dimensional} nature of images from camera systems; to make timely decisions, a monitoring approach needs to process such high-dimensional images in a \textit{scalable} fashion. The second Challenge \hypertarget{iilink}{}\iilink is that these abrupt changes typically involve \textit{structural} image features, e.g., changes in edges and/or shapes. The quick detection of such structural changes in high-dimensional images can thus be a difficult task. Finally, the third Challenge \hypertarget{iiilink}{}\iiilink is that a reliable and probabilistic quantification of \textit{uncertainty} is needed for confident change detection. Such uncertainty quantification (UQ) permits the assessment of risk in consequential downstream decisions, e.g., the costly interruption of manufacturing processes for maintenance. This need for scalability, structure-awareness and uncertainty quantification in change-point detection arises in broad scientific and engineering applications, which we elaborate on later.

% Such uncertainty is important not only for quantifying threat risk in surveillance applications, but also for establishing confidence in potential findings. 
% \cite{detectAbruptChange93} focuses on the design and investigation of online change detection algorithms. 

Despite its importance, there is scant literature on online change-point detection methods that jointly address these three challenges. Here, a change-point is defined as a sudden change in the data distribution over time. Much of the work in this literature (see \cite{poor-hadj-QCD-book-2008} for an overview) focuses on the cumulative sum (CUSUM) approach \cite{page-biometrica-1954}, which is widely employed for detecting distributional shifts \citep{Siegmund1985}. The Hotelling's $T^2$ approach \citep{hotelling1947multivariate} offers a natural extension for multivariate data. \cite{Lorden1971} explores a variety of detection stopping rules that achieve asymptotic optimality. \citep{tartakovsky2014sequential} provides further theoretical developments on sequential hypothesis testing and quickest change-point detection. Recent approaches investigate the use of kernel methods \cite{mmd,li2015m} and topological structure \cite{zheng2023percept}. Such developments, however, are largely frequentist in nature; given our need for reliable uncertainty quantification, a Bayesian approach may be more desirable here. An early work on Bayesian online change-point detection is \cite{Bayes_CUSUM_West}, which proposes a Bayesian CUSUM approach that leverages cumulative Bayes' factors for sequential change detection. This is further extended in \cite{Fearnhead_Liu} and \cite{BOCD}, which perform monitoring via the posterior distribution of its \textit{run length}, i.e., the time since its last change-point. Recent developments include \cite{pmlr-Knoblauch}, which explores the notion of model selection for change detection in spatiotemporal models; \cite{turner2009adaptive}, which extends this procedure via online hyperparameter learning; and \cite{Garnett2009SequentialBP, Saatci2010GaussianPC}, which explores the use of Gaussian process models. 

The above literature, however, largely focuses on lower-dimensional data, and may thus scale poorly for high-dimensional image data (see Challenge \ilink). Recent approaches have been explored for this challenging image setting. This includes \cite{Okhrin2021}, which uses generalized likelihood ratios on local image characteristics for computational scalability. \cite{Parallelized_Otto2021} explores the use of parallelized multivariate exponentially weighted moving average control charts, which leverages a careful segmentation of the image data. \cite{Yan_decomp2018} utilizes the decomposition of image data streams into functional means and sparse anomalies, then applies sequential likelihood ratio tests for anomaly monitoring. Such methods, however, do not provide a measure of monitoring uncertainty (see Challenge \iiilink). They also largely do not incorporate prior knowledge of image structure, e.g., edges or shapes, to guide change detection (see Challenge \iilink). As we shall see in later applications, neglecting such structure can considerably delay the online detection of changes.

Much work on integrating image structure for change detection (Challenge \iilink) arises in the image processing literature. To model such structure, these methods largely employ variations of convolutional neural networks (CNNs); see \citep{He2015DeepRL, Kiela2014LearningIE}. Given the large body of work on CNNs in machine learning \citep{NIPS2012_c399862d, Wang2020_highResolution}, such CNN-based approaches are highly scalable for large images, thus addressing Challenge \ilink. \cite{Sultani2018RealWorldSV} employs a deep multiple-instance ranking framework for anomaly detection. \cite{Luo2017RememberingHW} makes use of a trained CNN for appearance encoding with a convolutional long-short-term-memory motion model for detecting anomalies. A recent work \cite{MONAD_Doshi2021} explores a multi-objective deep-learning-based modeling framework for online anomaly monitoring. Such approaches, however, do not offer the desired quantification of monitoring uncertainty (Challenge \iiilink) that naturally comes from a Bayesian framework. Furthermore, for such deep models, this lack of UQ may inflate one's confidence of the fitted model, which may result in erratic detection performance with many false alarms; we shall see this later.

To jointly address Challenges \ilink-\iiilink, we propose a new Bayesian onLine Structure-Aware change deTection (BLAST) procedure. BLAST first leverages the deep Gaussian Markov Random Field (GMRF) model proposed in \cite{DGMRF}, to elicit expected image structure from offline reference images as prior knowledge. This deep GMRF offers two nice advantages. First, it enjoys much of the same scalability as GMRFs, which are widely used for large-scale modeling in spatial statistics; see, e.g., \cite{hartman2008fast, SIMPSON201216, Nguyen2016_sensor, mak2016regional}. Second, it can be viewed as a probabilistic multi-layer CNN \citep{DGMRF}, and thus provides a flexible framework for eliciting image structure. Using this elicited deep GMRF prior, BLAST then employs a new Bayesian online procedure for image monitoring via the so-called posterior run length distribution in \cite{BOCD}. This posterior run length distribution can be computed in an online fashion using $\mathcal{O}(p^2)$ work at each time-step, where $p$ is the number of image pixels. This permits scalable monitoring and efficient detection of changes in high-dimensional images, while also providing the desired quantification of Bayesian uncertainty. We compare BLAST with the state-of-the-art in a suite of numerical experiments, and explore its effectiveness in two applications, the first on street scene monitoring and the second on real-time process monitoring for metal additive manufacturing.

This paper is organized as follows. Section 2 provides an overview of existing methods, and investigates their potential limitations in our street scene monitoring application. Sections 3 and 4 present our BLAST framework: Section 3 outlines its prior elicitation procedure via the deep GMRF, and Section 4 proposes a scalable procedure for efficient online updates of the posterior run length distribution. Section 5 explores the performance of BLAST compared to the state-of-the-art in a suite of numerical experiments. Section 6 demonstrates the effectiveness of BLAST in two applications, the first on street scene monitoring and the second on process monitoring for metal additive manufacturing. Section 7 concludes the paper.

\section{Background and Motivation}

We first provide a brief overview of several key baseline methods: the Hotelling's $T^2$ method \cite{hotelling1947multivariate}, the maximum mean discrepancy (MMD) approach \cite{mmd}, and the Multi-Objective Neural Anomaly Detector (MONAD) method \cite{MONAD_Doshi2021}. We then investigate potential limitations of these methods in our motivating street scene monitoring application.

\subsection{Existing Baseline Methods}
\label{sec:existing_methods}

The first baseline monitoring approach is the classic Hotelling's-$T^2$ statistic \citep{hotelling1947multivariate}, which is widely used for monitoring multivariate data in engineering applications \citep{xue2006model,boullosa2017monitoring}. Let $\mathbf{x}_t \in \mathbb{R}^p$ be the multivariate data observed at time $t$. Suppose its ``pre-change'' mean vector and covariance matrix, denoted $\boldsymbol{\mu}_{\rm pre}$ and $\boldsymbol{\Sigma}_{\rm pre}$, can be estimated well from offline reference data. Denote $S_t^{\rm H}$ as the Hotelling's-$T^2$ monitoring statistic at time $t$. With $S_1^{\rm H} = 0$ at time $1$, this monitoring statistic is defined recursively for times $t = 2, 3, \cdots$ as:
% %\vspace{-0.2in}
% \begin{equation}
% \setlength{\abovedisplayskip}{5pt}
% \setlength{\belowdisplayskip}{5pt}
%     (\bar{\mathbf{x}} - \boldsymbol{\mu}_0)^\intercal \Sigma_0^{-1} (\bar{\mathbf{x}} - \boldsymbol{\mu}_0),
% \end{equation}
% %\vspace{-0.2in}
% \noindent where $\bar{\mathbf{x}}$ is the sample mean of the vevtor 
% %
% The vanilla Hotelling's $T^2$ statistic is calculated using \textit{only} data at the current time $t$, with all historical data discarded. 
% To compute the test statistic more efficiently, it can be coupled with the CUSUM scheme \citep{page-biometrica-1954}, which makes use of a cumulative sum of the statistic over time. The resulting detection statistic $S_t^H$ is then given by: 
%\vspace{-0.2in}
\begin{equation}
\setlength{\abovedisplayskip}{5pt}
\setlength{\belowdisplayskip}{5pt}
S_t^{\rm H} = (S_{t-1}^{\rm H})^{+} + (\bar{\mathbf{x}}_{(t-w):t} - \boldsymbol{\mu}_{\rm pre} )^\intercal \boldsymbol{\Sigma}_{\rm pre}^{-1} (\bar{\mathbf{x}}_{(t-w):t} - {\boldsymbol{\mu}_{\rm pre}} ) - d^{\rm H}.
    \label{eq:hotelling}
\end{equation}
%\vspace{-0.1in}
Here, $(x)^+ = \max(x, 0)$ returns the non-negative part of $x$, and $\bar{\mathbf{x}}_{(t-w):t}$ denotes the sample average of the data vectors $\{\mathbf{x}_{t-w},\ldots,\mathbf{x}_{t}\}$. The constant $d^{\rm H}$ is a drift parameter that is typically estimated from historical reference data \cite{zheng2023percept}. The specific form of \eqref{eq:hotelling} integrates a cumulative sum scheme (CUSUM; \cite{page-biometrica-1954}), which accelerates the detection of potential changes; such a form is widely adopted in the literature \citep{zheng2023percept}. A change is then declared when the monitoring statistic $S_t^{\rm H}$ exceeds a pre-specified threshold.

In our set-up, however, the observed data at time $t$ takes the form of an image, which we denote by the matrix $\mathbf{X}_t \in \mathbb{R}^{q_1 \times q_2}$. A natural approach is to first vectorize this as $\mathbf{x}_t = \text{vec}(\mathbf{X}_t)$, then apply the Hotelling-$T^2$ statistic \eqref{eq:hotelling}. To deal with the high-dimensional nature of $\mathbf{x}_t$, one can further perform on such vectors principal component analysis (PCA; \citep{pearson1901liii}), then use its corresponding weights as the data vectors within the monitoring statistic \eqref{eq:hotelling}. Projections for such principal components can be estimated from historical pre-change data; details on such a procedure can be found in \cite{xie2021sequential,zheng2023percept}. 

% When the data is known to be concentrated on a linear subspace, one can adapt the Hotelling's $T^2$ test by first performing Principal Component Analysis (PCA), then using the resulting principal components as data within equation \eqref{eq:hotelling}. With this, a change-point is then declared at time $t$ when the statistic $S_t^{H}$ first exceeds a pre-specified threshold $b$.

%\cmtS{a paragraph on MMD monitoring?}

A potential limitation of the Hotelling-$T^2$ is that it is highly parametric in nature. The second baseline method offers a nonparametric alternative via the maximum mean discrepancy (MMD) test \citep{mmd,li2015m}. Suppose we are at time $t$, and wish to investigate whether a change-point occurred at time $\xi < t$, where $\xi > \lceil t/2 \rceil$. With $n = t-\xi$, let $\{\mathbf{x}_{{\rm pre},i}\}_{i=1}^n$ and $\{\mathbf{x}_{{\rm post},i}\}_{i=1}^n$ be the observed data vectors at the pre-change times $\{\xi-n+1,\cdots, \xi\}$ and the post-change times $\{\xi+1, \cdots, t\}$, respectively. With this, consider the following test statistic:
\begin{equation}
\setlength{\abovedisplayskip}{10pt}
\setlength{\belowdisplayskip}{10pt}
\rho_{t,\xi} = \frac{1}{n^2} \sum_{i,i'=1}^{n} k(\mathbf{x}_{{\rm pre},i}, \mathbf{x}_{{\rm pre},i'}) + \frac{1}{n^2} \sum_{j,j'=1}^{n} k(\mathbf{x}_{{\rm post},j}, \mathbf{x}_{{\rm post},j'}) - \frac{2}{n^2} \sum_{i=1}^{n}\sum_{j=1}^{n} k(\mathbf{x}_{{\rm pre},i}, \mathbf{x}_{{\rm post},j}),
\label{eq:mmd}
\end{equation}
where $k$ is a chosen symmetric positive-definite kernel, e.g., the squared-exponential kernel. The test statistic $\xi_{t,\xi}$ provides a nonparametric test of distributional equivalence between the pre- and post-change data. To investigate whether a change-point occurred at \textit{any} time prior to $t$, the MMD statistic takes the maximum statistic of $\rho_{t,\xi}$ over $\xi$, i.e.:
\begin{equation}
S_t^{\rm MMD} = \max_{\xi < t, \xi > \lceil t/2 \rceil} \rho_{t,\xi}.
\label{eq:mmdstat}
\end{equation}
As before, for image data, one can first vectorize the images $\mathbf{x}_t=\text{vec}(\mathbf{X}_t)$ prior to monitoring. A change is similarly declared when the monitoring statistic $S_t^{\rm MMD}$ exceeds a pre-specified threshold.

% Given a class of functions $\mathcal F$ and two distributions $p$ and $q$, the MMD distance between $p$ and $q$ is defined as 
% $\mathrm{MMD}_{\mathcal F}(p,q) = \sup_{f\in\mathcal F}(\mathbb E_{x\sim p}[f(x)] - \mathbb E_{y\sim q}[f(y)])$. 
% When $\mathcal{F}$ is a reproducing kernel Hilbert space (RKHS) associated with kernel function $K(\cdot,\cdot)$, this MMD statistic can be written as: 

One drawback with the above two baselines is that they do not leverage the underlying image structure (e.g., shapes, edges) in $\mathbf{X}_t$ for quick change detection. As we show later, when such structure is present in the data, ignoring it can considerable delays in online change detection. The third baseline, called the Multi-Objective Neural Anomaly Detector (MONAD; \cite{MONAD_Doshi2021}), offers a solution; this is a recent baseline from the literature on CNN-based approaches. MONAD follows two steps. The first step employs a generative adversarial network \cite{Unet} for future frame prediction, along with a deep-learning-based object detector \cite{YOLO} for image feature extraction. The second step uses the extracted features within a CUSUM-like framework for online change detection. For brevity, we simply refer to the MONAD test statistic as $S_t^{\rm MONAD}$, and refer the reader to \cite{MONAD_Doshi2021} for further details on this procedure.

% deep learning-based features, consisting of the future frame predicting error using a Generative Adversarial Network (GAN) \cite{Unet}, and location and appearance features using an object detection system called You Only Look Once (YOLO) \cite{YOLO}. The second module is to use the extracted features to construct a monitoring test statistic using the $k-$NN (Euclidean) distance to the nominal space learned from training data. The proposed anomaly detection system computes the frame-level anomaly evidence $\delta_t$:
% \begin{equation}
% \delta_t = \big(\max_i\{d_t^i\}\big)^m - \big(d_\alpha \big)^m,
% \end{equation}
% where $m$ is the dimensionality of feature vector $F_t^i$ for the detected object $i$, and $d_\alpha$ is the drift parameter that can also be trained using historical data under significance level $\alpha$. Following a CUSUM-like procedure \cite{page-biometrica-1954}, the test statistic $S_t^{M}$ is defined as 
% \begin{equation}
% S_t^{M} = \max\{S_{t-1}^{M} + \delta_t, 0\}, \quad S_0^{M} = 0.
% \end{equation}
% One then declares a change-point at time $t$ when the test statistic $S_t^{M}$ exceeds some pre-specified threshold $h$.

\subsection{Street Scene Monitoring}
\label{sec:mot}

% Surveillance cameras are ubiquitous, and the automation of anomalous activities can greatly reduce human effort. In most cases, almost all of the video from a surveillance camera is unimportant and only unusual video segments are of interest.

\begin{figure}[!t]
%\if1\fast
%{

\centering
    \centering
    % \begin{tabular}{c}
    %  \includegraphics[width=0.9\textwidth]{Car/data_visual.png}  \\
    %  (a)
    %  \end{tabular}
    %  \vspace{0.1in}
     
    %  \begin{tabular}{ccc}
    %      \includegraphics[width=.3\textwidth]{car_Hotelling.png} &
    %      \includegraphics[width=.3\textwidth]{Car/car_mmd.png} &
    %      \includegraphics[width=.3\textwidth]{car_MONAD.png} \\
    %      (b) & (c) & (d)
    %  \end{tabular}
    \includegraphics[width=0.9\textwidth]{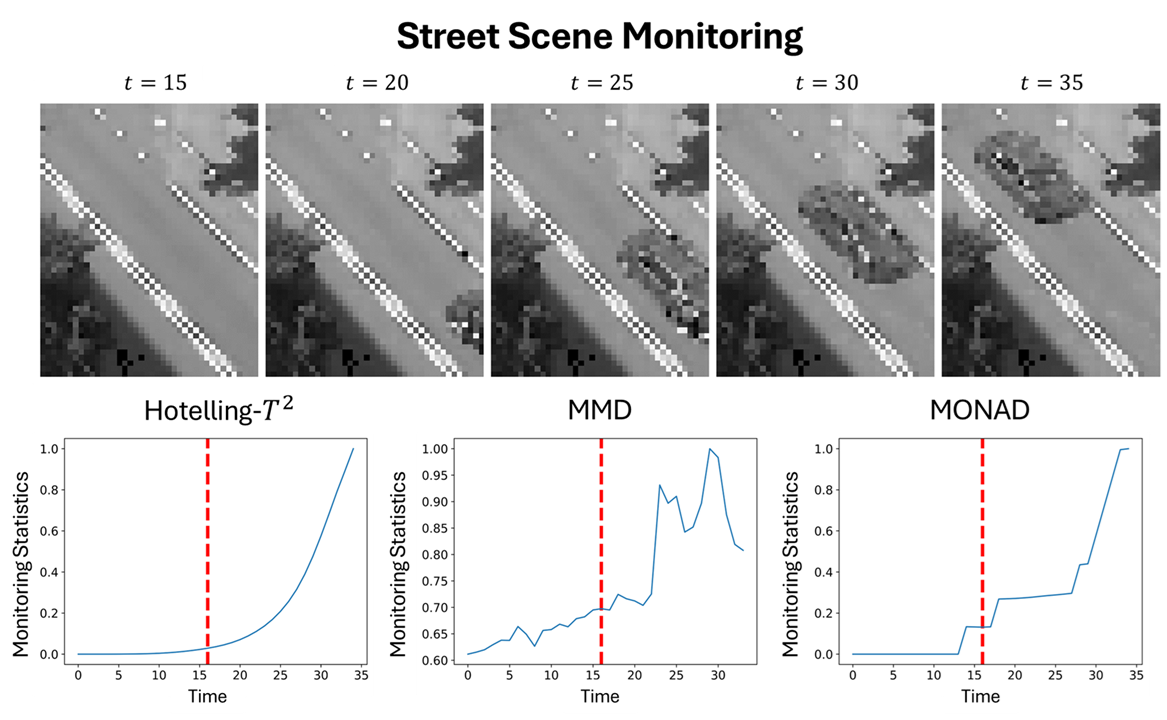}
     \caption{(Top) Image snapshots of a car passing through a street scene. Here, the car arrives at time $t^*=16$, and continues to pass by after. (Bottom) Monitoring statistics of existing methods for the street scene application, with the true change-point $t^*=16$ dotted in red.}
     \vspace{-0.1in}
        \label{fig:motivation}
\end{figure}

We investigate next the performance of these baseline methods for a street scene monitoring application, which we later revisit in Section \ref{sec:street}. The set-up here is taken from \cite{car_data}: $T=35$ image snapshots (each with 50 $\times$ 40 pixels) are taken from a static USB camera, looking down on a scene of a two-lane street with bike lanes and pedestrian sidewalks. The change-point we wish to detect is a car entering the scene at time $t^*=16$. Figure \ref{fig:motivation} (top) shows several snapshots of the street scene before and after this change-point. These images can be seen to have highly structured image features, e.g., trees and road markings, that can potentially be leveraged for quick change detection (see Challenge \iilink).

We now apply the aforementioned baseline methods. As discussed earlier, the Hotelling-$T^2$ is implemented by first vectorizing the image data $\mathbf{X}_t$, then applying PCA on the vectorized data and taking the top 15 principal component weights as the data $\mathbf{x}_t$ within the monitoring statistic \eqref{eq:hotelling}. The MMD test is similarly implemented by vectorizing the image data. Here, the kernel $k$ is taken as the squared-exponential kernel $k(\mathbf{x},\mathbf{x}')=\exp\{-\theta\|\mathbf{x}-\mathbf{x}'\|_2^2\}$, where $\theta$ is set following the specification in \cite{scholkopf2002learning}. Finally, MONAD is implemented using code provided by the authors \cite{MONAD_Doshi2021}.

Figure \ref{fig:motivation} (bottom) show the monitoring statistics as a function of time $t$ for the three baselines. For the Hotelling-$T^2$, we see that its monitoring statistics increase at a gradual rate after the change-point at $t^*=16$. Such an approach can thus be quite slow in identifying the desired change, which is not surprising in retrospect as it largely ignores structural features in the underlying image data (see Challenge \iilink). For the MMD, we observe a similar phenomenon of delayed detection; this is again not surprising since the MMD also largely ignores structural image features. Finally, for MONAD (which leverages image features learned via a deep-learning-based framework), we see noticeable instabilities for its monitoring statistics: prior to the true change at $t^*=16$, its monitoring statistics rise considerably, which raises the risk of detection false alarms. A likely reason for this is the aforementioned lack of uncertainty quantification (see Challenge \iiilink) on its learned image features. Neglecting such uncertainties inflates one's confidence in the learning model, which may result in frequent false alarms in monitoring. To tackle these challenges, we thus want a scalable \textit{Bayesian} change-point approach, which can leverage a \textit{structure-aware} and \textit{probabilistic} image model to guide a confident and quick detection of changes; this is introduced next.

%The larger the resolution gap between the training images and the test images, the more challenging it might be for the CNN to perform well on the low-resolution images. This is because significant details that the model learned to recognize in high-resolution images may not be as discernible in low-resolution images.
%CNNs learn hierarchical features, with early layers typically learning fine details (edges, textures) and deeper layers learning more abstract and global features (object parts, entire objects). If a model is trained on high-resolution images, its filters may be tuned to features that could be less visible or absent in lower-resolution images. The effective receptive field of the CNN might not align well with the scale of features in low-resolution images, potentially affecting performance.

\section{BLAST: Image Prior Elicitation}
\label{sec:prior}

\begin{figure}[!t]
    \centering
    \includegraphics[width = \textwidth]{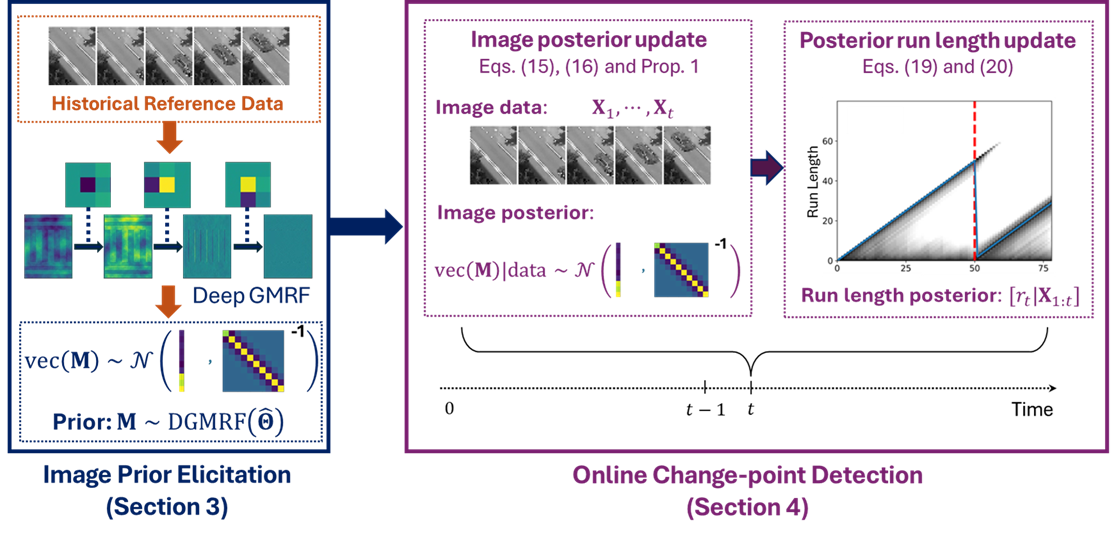}
    \caption{A workflow outlining the proposed BLAST method for probabilistic, scalable and structure-aware online change-point detection with image data.}
    \label{fig:framework}
\end{figure}

We present in the following our BLAST framework, which targets the aforementioned three challenges. We first outline in this section the employed prior elicitation procedure for capturing desired image structure, using the deep Gaussian Markov random field in \cite{DGMRF}. We then show in the next section how this elicited prior can be leveraged for scalable and structure-aware online change-point detection with image data. Figure \ref{fig:framework} visualizes the workflow for the proposed BLAST framework.

% and its capacity for eliciting image structure within a probabilistic model. We then show that this deep GMRF nicely integrates within an online Bayesian change-point framework, allowing for efficient probabilistic monitoring via scalable computation of its posterior run length

\subsection{Deep Gaussian Markov Random Fields}
\label{sec:dgmrf}

Consider a random image $\mathbf{X} \in \mathbb{R}^{q_1 \times q_2}$, and let $\mathbf{x} = \text{vec}(\mathbf{X}) \in \mathbb{R}^p$ be its vectorized representation. Let $\mathcal{G} = (\mathcal{V},\mathcal{E})$ be a graph where $\mathcal{V} = \{1, \cdots, p\}$ is the set of indices and $\mathcal{E}$ is the set of corresponding edges. For two indices $i$ and $j$ in $\mathcal{V}$, let $i$ and $j$ be connected (denoted $i \sim j$) if the edge $(i,j)$ exists in $\mathcal{E}$. With this, $\mathbf{x}$ follows a Gaussian Markov random field \citep{banerjee2003hierarchical} if there exists a graph $\mathcal{G}$ such that: 
\begin{equation}
\mathbf{x} \sim \mathcal{N}(\boldsymbol{\mu}, \mathbf{Q}^{-1}),
\label{eq:gmrf}
\end{equation}
where $\boldsymbol{\mu}$ is its mean vector, and $\mathbf{Q}$ is its precision matrix with $(i,j)$-th entry $Q_{i,j} = 0$ whenever $i$ is not connected to $j$ in $\mathcal{G}$. For the $i$-th entry of $\mathbf{x}$, i.e., $x_i$, this ensures its conditional distribution $x_i|\{x_j:j \neq i\}$ equals $x_i|\{x_j: j \sim i\}$. In other words, the conditional distribution of an image pixel given all other pixels depends only on its connected pixels in $\mathcal{G}$. This ``local'' dependence of GMRFs, coupled with a sparse representation for $\mathcal{G}$, has been widely leveraged for scalable probabilistic modeling in spatial statistics \cite{lindgren2011explicit,mak2016regional} and computer experiments \cite{ding2019bdrygp}.

% $\mathbf{Q}$ should further satisfy: For each GMRF, there exists a graph with vertices $\mathcal{V}$ corresponding to the elements in $\mathbf{x}$, and edges $\mathcal{E}$ define the conditional independencies of elements. Given the neighbors of an element in $\mathbf{x}$, the element is conditionally independent of all other elements in $\mathbf{x}$. Mathematically, every GMRF has the property
% \[{i,j} \in \mathcal{E} \longleftrightarrow Q_{i,j} \not = 0, \quad \text{for all } i \not = j,\]
% and resulting in a sparse precision matrix $\mathbf{Q}$. This is one key property of GMRFs, and the "local" dependence structure can be used to simplify computations, especially in high-dimensional problems.

\begin{figure}[!t]
    \centering
    \includegraphics[width = \textwidth]{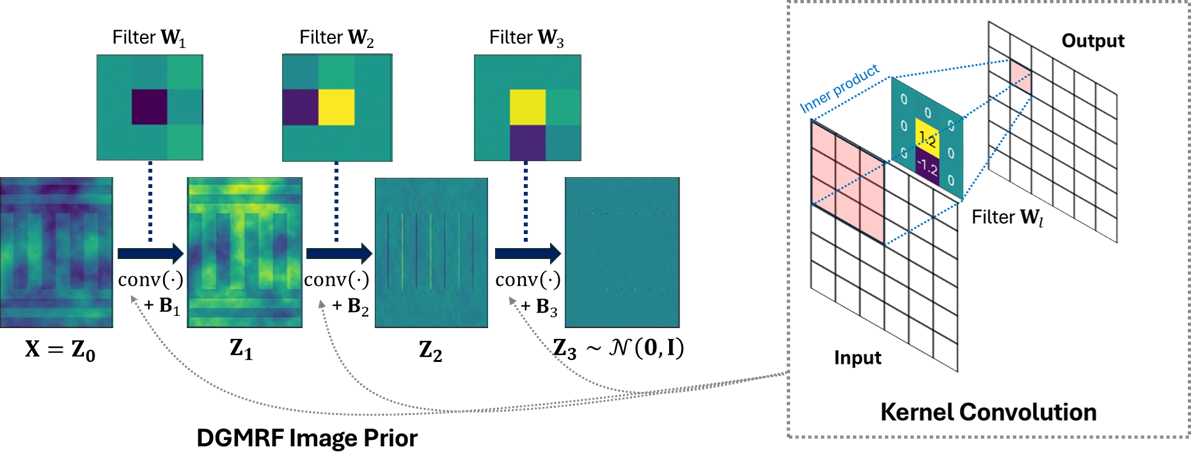}
    \caption{Visualizing the DGMRF prior in \cite{DGMRF}. [Left] Visualizing the convolutional construction \eqref{eq:conv} for a $L=3$-layer DGMRF. Here, the filters $\mathbf{W}_l$ and intercept matrices $\mathbf{B}_l$ are elicited via a reference image dataset with vertical and horizontal edge structure. [Right] Visualizing the kernel convolution operation $\textup{conv}(\cdot,\mathbf{W}_l)$ in Equation \eqref{eq:conv}.}
    \label{fig:deepGMRF}
\end{figure}

The deep GMRF in \cite{DGMRF} adapts this GMRF for probabilistic image modeling. This model uses the following $L$-layer convolutional neural network construction, where $L>1$. Let $\mathbf{Z}_0 = \mathbf{X}$ be the modeled image of interest, and let $\mathbf{Z}_l \in \mathbb{R}^{q_1 \times q_2}$ denote the image in the $l$-th CNN layer. Suppose each layer follows the convolutional construction:
\begin{equation}
\mathbf{Z}_l = \text{conv}(\mathbf{Z}_{l-1},\mathbf{W}_l) + \mathbf{B}_l, \quad l = 1, \cdots, L,
\label{eq:conv}
\end{equation}
where $\text{conv}(\cdot)$ is the convolution operator; see Figure \ref{fig:deepGMRF} (right). Here, $\mathbf{W}_l \in \mathbb{R}^{3 \times 3}$ serves as the filter matrix for image convolution in the $l$-th layer\footnote{While larger filter matrices (i.e., with dimensions greater than $3 \times 3$) can be used, we elect for $3 \times 3$ filters here for computational efficiency in determinant and inverse computations, as discussed later.}, and $\mathbf{B}_l \in \mathbb{R}^{q_1 \times q_2}$ is its intercept matrix. One can show that \eqref{eq:conv} induces a linear map $g_l$ on its vectorized components $\mathbf{z}_l = \text{vec}(\mathbf{Z}_l)$, given by:
\begin{equation}
\mathbf{z}_l = g_l(\mathbf{z}_{l-1}) := \mathbf{G}_l \mathbf{z}_{l-1} + \mathbf{b}_l,
\end{equation}
where $\mathbf{b}_l = \text{vec}(\mathbf{B}_{l})$, and $\mathbf{G}_l \in \mathbb{R}^{p \times p}$ is an invertible matrix that can be constructed from filter $\mathbf{W}_l$ (see \cite{DGMRF} for details). Here, the primary reason for defining $\mathbf{X}$ via this sequence of \textit{inverse} transforms is to formalize the following connection to GMRFs; since $\mathbf{G}_l$ is invertible, one can define an analogous sequence of forward transforms to define $\mathbf{X}$.

% In later experiments, we consider filters of sizes $3 \times 3$, as suggested in \cite{DGMRF}.

With this convolutional construction, \cite{DGMRF} shows that the modeled image $\mathbf{X}$ resembles a GMRF under a certain specification. First, suppose the last image layer $\mathbf{Z}_L$ follows a standard normal distribution $\text{vec}(\mathbf{Z}_L) \sim \mathcal{N}(\mathbf{0},\mathbf{I})$. Next, for each layer $l$, suppose the filter matrix $\mathbf{W}_l$ follows one of two forms:
\begin{equation}
\mathbf{W}_l^{\rm +} = \begin{bmatrix}
0 & a_{l,3} & 0 \\
a_{l,2} & a_{l,1} & a_{l,4}\\
0 & a_{l,5} & 0
\end{bmatrix} \quad \text{or} \quad \mathbf{W}_l^{\rm seq} = \begin{bmatrix}
0 & 0 & 0 \\
0 & a_{l,1} & a_{l,2}\\
a_{l,3} & a_{l,4} & a_{l,5}
\end{bmatrix}.
\label{eq:filter}
\end{equation}
Such filters are known as ``+'' and ``seq'' filters in \cite{DGMRF}; its rationale is discussed later. With this, \cite{DGMRF} shows that the vectorized image $\text{vec}(\mathbf{X})$ follows a GMRF of the form:
\begin{equation}
\text{vec}(\mathbf{X}) \sim \mathcal{N}(\boldsymbol{\mu}_{\mathbf{G}},\mathbf{Q}_{\mathbf{G}}^{-1}),
\label{eq:gmrfexpn}
\end{equation}
where its mean vector and covariance matrix are given by $\boldsymbol{\mu}_{\mathbf{G}} = -\mathbf{G}^{-1}\mathbf{b}$ and $\mathbf{Q}_{\mathbf{G}}^{-1} = (\mathbf{G}^T\mathbf{G})^{-1}$ respectively, with $\mathbf{G} = \mathbf{G}_L \mathbf{G}_{L-1} \cdots \mathbf{G}_1$ and $\mathbf{b} = g(\bm{0}) := [g_L \circ g_{L-1} \circ \cdots g_1]( \bm{0})$. Furthermore, with sparse filters on $\mathbf{W}_l$ (e.g., the two filters in \eqref{eq:filter}), the precision matrix $\mathbf{Q}_{\mathbf{G}}$ of this GMRF can be shown to be sparse as well \citep{DGMRF}. We denote the above Bayesian model on image $\mathbf{X}$ as $\mathbf{X} \sim \text{DGMRF}(\boldsymbol{\Theta})$, where $\boldsymbol{\Theta} = \{(\{a_{l,j}\}_{j=1}^5,\mathbf{b}_l)\}_{l=1}^L$ denote its \textit{model} parameters, i.e., its filter weights and intercepts at each layer.

% The above link thus provides a nice way of constructing certain types of GMRFs via a sequence of image convolutions. 

% $\mathbf{Q}_{\mathbf{G}}$ can further be shown to satisfy the desired adjacency condition $i \sim j \Leftrightarrow Q_{i,j} = 0$, thus the vectorized image from this deep model follows a GMRF.

For our use later in change-point detection, the above convolutional construction (and its link to a GMRF) provides a nice prior image model that jointly (i) captures expected image structure, and (ii) facilitates scalable computation for efficient online change detection. We explore next point (i) by showing how $\boldsymbol{\Theta}$ can be specified from historical pre-change data, such that the deep GMRF prior $\text{DGMRF}(\boldsymbol{\Theta})$ captures desirable image features. We then investigate in Section \ref{sec:blast} how this elicited prior can be used for point (ii), i.e., the efficient detection of online changes in image data.

% \subsection{Deep Gaussian Markov Random Fields}
% \label{DGMRF}
\subsection{Prior Elicitation from Historical Reference Data}
\label{sec:prior_elicitation}

Now, suppose we have offline reference image data of the form $\tilde{\mathbf{X}}_1, \cdots, \tilde{\mathbf{X}}_N \in \mathbb{R}^{q_1 \times q_2}$. If we expect similar image features to be present in the online detection setting, then such data can be used to provide an informed elicitation of $\boldsymbol{\Theta}$ for the DGMRF prior. This can be efficiently performed via an adaptation of the variational inference approach in \cite{DGMRF}, which we describe below.

Suppose the reference data $\tilde{\mathbf{X}}_1, \cdots, \tilde{\mathbf{X}}_N$ are sampled with i.i.d. noise $\mathcal{N}(0,\gamma^2)$ from the underlying images $\mathbf{M}_1, \cdots, \mathbf{M}_N$. Further suppose $\mathbf{M}_1, \cdots, \mathbf{M}_N$ independently follow the same $\text{DGMRF}(\boldsymbol{\Theta})$ prior, i.e., they share similar image features modeled by the image filters in $\boldsymbol{\Theta}$. The marginal likelihood of $\tilde{\mathbf{X}}_1, \cdots, \tilde{\mathbf{X}}_N$ can then be written as\footnote{Here, $[X]$ denotes the ``distribution of'' random variable $X$.}:
\begin{equation}
[\tilde{\mathbf{X}}_1, \cdots, \tilde{\mathbf{X}}_N|\boldsymbol{\Theta}] = \frac{\prod_{j=1}^N [\tilde{\mathbf{X}}_j|\mathbf{M}_j,\boldsymbol{\Theta}][\mathbf{M}_j|\boldsymbol{\Theta}]}{\prod_{j=1}^N [\mathbf{M}_j|\tilde{\mathbf{X}}_j,\boldsymbol{\Theta}]},
\label{eq:marglkhd}
\end{equation}
for an arbitrary choice of $\mathbf{M}_1, \cdots, \mathbf{M}_N$. One way to elicit $\boldsymbol{\Theta}$ is via empirical Bayes, which maximizes the marginal likelihood \eqref{eq:marglkhd} with respect to $\boldsymbol{\Theta}$. A bottleneck with this approach, however, is that each evaluation of \eqref{eq:marglkhd} requires the determinant of a $p \times p$ precision matrix, which incurs $\mathcal{O}(p^3)$ work and can thus be unwieldy for large images.

To address this, we make use of a variational evidence lower bound (ELBO; \cite{Blei2017}) approach that lower bounds the above marginal likelihood. This extends the approach in \cite{DGMRF} for the current set-up of multiple (i.e., $N>1$) reference images $\tilde{\mathbf{X}}_{1:N} := \{\tilde{\mathbf{X}}_1, \cdots, \tilde{\mathbf{X}}_N\}$. For the above model, this ELBO takes the form: 
\begin{equation}
\text{ELBO}(\boldsymbol{\Theta},\boldsymbol{\Phi}, \gamma^2;\tilde{\mathbf{X}}_{1:N}) =  \mathbb{E}_{q_{\boldsymbol{\Phi}}(\mathbf{M})} \left\{-\log q_{\boldsymbol{\Phi}}(\mathbf{M}) + \sum_{j=1}^N \log [\tilde{\mathbf{X}}_j,\mathbf{M}_j|\boldsymbol{\Theta}] \right\},
\label{eq:elbo}
\end{equation}
where $q_{\boldsymbol{\Phi}}(\mathbf{M})$ is the variational posterior approximation for $\mathbf{M} = (\mathbf{M}_1, \cdots, \mathbf{M}_N)$ depending on \textit{variational} parameters $\boldsymbol{\Phi}$. We adopt for $q_{\boldsymbol{\Phi}}(\mathbf{M})$ the Gaussian variational family $q_{\boldsymbol{\Phi}}(\mathbf{M}) = \prod_{j=1}^N \mathcal{N}(\text{vec}(\mathbf{M}_j);\boldsymbol{\nu}_{\boldsymbol{\Phi}},\mathbf{S}_{\boldsymbol{\Phi}})$, with mean vector $\boldsymbol{\nu}_{\boldsymbol{\Phi}}$ and diagonal covariance matrix $\mathbf{S}_{\boldsymbol{\Phi}}$. With this, the ELBO can be simplified as:
\begin{align}
\begin{split}
\text{ELBO}(\boldsymbol{\Theta},\boldsymbol{\Phi}, \gamma^2;\tilde{\mathbf{X}}_{1:N}) = & \sum_{j=1}^{N} \Bigg\{\frac{1}{2} \log \det \mathbf{S}_{\boldsymbol{\Phi}} - p \log \gamma + \log \det \mathbf{G}_{\boldsymbol{\Theta}} - \\
& \quad \frac{1}{2} \mathbb{E}_{q_{\boldsymbol{\Phi}}(\mathbf{M})} \left[ g_{\boldsymbol{\Theta}}(\text{vec}(\mathbf{M}_j))^{\intercal} g_{\boldsymbol{\Theta}}(\text{vec}(\mathbf{M}_j))
  +{\frac{1}{\gamma^2}} \|\tilde{\mathbf{X}}_j - \mathbf{M}_j\|_F^2 \right] \Bigg\},
\label{eq:elbo2}
\end{split}
\end{align}
where constant terms have been omitted. Here, $g_{\boldsymbol{\Theta}}$, $\mathbf{G}_{\boldsymbol{\Theta}}$, $\boldsymbol{\mu}_{\boldsymbol{\Theta}}$ and $\mathbf{Q}_{\boldsymbol{\Theta}}$ follow from $g$, $\mathbf{G}$, $\boldsymbol{\mu}_{\mathbf{G}}$ and $\mathbf{Q}_{\mathbf{G}}$ from Section \ref{sec:dgmrf} using parameters $\boldsymbol{\Theta}$, respectively. One can then jointly optimize the ELBO \eqref{eq:elbo2} with respect to the model parameters $\boldsymbol{\Theta}$, variational parameters $\boldsymbol{\Phi}$ and noise variance $\gamma^2$, then use the optimized parameters $\hat{\boldsymbol{\Theta}}$ for the elicited prior $\text{DGMRF}(\hat{\boldsymbol{\Theta}})$ within the online change detection procedure in Section \ref{sec:blast}. Algorithm \ref{alg:alg1} summarizes these steps for image prior elicitation.

\begin{algorithm}[!t]
\caption{BLAST -- Image Prior Elicitation}\label{alg:alg1}
\begin{algorithmic}
\STATE \textit{\underline{Inputs}}: Historical image data $\tilde{\mathbf{X}}_1, \cdots, \tilde{\mathbf{X}}_N$, number of DGMRF layers $L$, filter choice (\texttt{+} or \texttt{seq}).
\STATE $\bullet$ \hspace{0.2cm} Initialize the model parameters $\boldsymbol{\Theta}_0$: filter weights and intercepts are initialized randomly from i.i.d. $\mathcal{N}(0,0.01)$ distributions. Initialize $\gamma_0^2$ as 1.0.

\STATE $\bullet$ \hspace{0.2cm} Initialize the variational parameters $\boldsymbol{\Phi}_0$: the mean vector $\boldsymbol{\nu}_{\boldsymbol{\Phi}}$ is initialized as the sample average of historical data, and entries of the diagonal matrix $\mathbf{S}_{\boldsymbol{\Phi}}$ are initialized randomly from i.i.d. truncated  $\mathcal{N}(0,1)$ distributions to ensure positivity.

\STATE $\bullet$ \hspace{0.2cm} Using initialized parameters, perform the optimization:
\[(\hat{\boldsymbol{\Theta}},\hat{\boldsymbol{\Phi}},\hat{\gamma^2}) \leftarrow \text{argmax}_{\boldsymbol{\Theta},\boldsymbol{\Phi},\gamma^2} \; \text{ELBO}(\boldsymbol{\Theta},\boldsymbol{\Phi},\gamma^2; \tilde{\mathbf{X}}_{1:N})\]
using the Adam optimizer \cite{Adam}. Here, the ELBO objective follows from Equation \eqref{eq:elbo2}.

\STATE $\bullet$ \hspace{0.2cm} Construct the elicited image prior $\text{DGMRF}(\hat{\boldsymbol{\Theta}})$ using Equation \eqref{eq:gmrfexpn}.
% values for the model parameters $\mathbf{\theta} \subset \{(\mathbf{w}_l, \mathbf{b}_l)_{l = 1, \cdots, L}, \sigma_2\}$ and the variational parameters $\mathbf{\phi} = \{\mathbf{\nu}, \mathbf{S}\}$

% to obtain $\hat{\mathbf{\theta}}$ and $\hat{\mathbf{\phi}}$.
% \STATE
\STATE \textit{\underline{Output}}: Elicited image prior $\text{DGMRF}(\hat{\boldsymbol{\Theta}})$.
\end{algorithmic}
\end{algorithm}

There are two key reasons why the ELBO form \eqref{eq:elbo2} facilitates efficient optimization and thus scalable prior elicitation. First, using either the + or seq filters from \eqref{eq:filter}, the log-determinant term $\log \det \mathbf{G}_{\boldsymbol{\Theta}}$ can be efficiently computed in $\mathcal{O}(p)$ work (see Sections 3.2.1 and 3.2.2 of \cite{DGMRF}). Second, gradient estimates of the ELBO \eqref{eq:elbo2} can be obtained via Monte Carlo sampling from $q_{\boldsymbol{\Phi}}(\mathbf{M})$, which integrates directly within state-of-the-art stochastic gradient optimizers with automatic differentiation \cite{baydin2018automatic} and backpropagation \cite{chauvin2013backpropagation}. In our later implementation, we made use of the popular Adam optimizer \cite{Adam} from the \texttt{PyTorch} library \citep{paszke2019pytorch} to perform this optimization for prior elicitation.

There are similarly two key advantages in using  + or seq filters in \eqref{eq:filter}. The first is its capacity for learning desirable image structures: + filters can capture edge, pattern or texture structure, whereas seq filters can capture sequential patterns with neighboring pixels \cite{DGMRF}. Using such filters with carefully elicited weights (via the above variational procedure), one can capture expected image features that can be leveraged for quick change detection. To see this, consider the following illustrative example. Here, $N=30$ reference images are generated with vertical and horizontal edge structure (see Figure \ref{fig:deepGMRF} left), with its underlying spatial structure sampled from a Mat\'ern GMRF \cite{lindgren2011explicit}; details on this in Section \ref{sec:spatchange}. With this reference data, we apply the earlier variational procedure to fit DGMRF filter weights, using $L=3$ layers and + filters. Figure \ref{fig:deepGMRF} (top left) shows the fitted filters for each of the three layers. We see that the last two filters $\hat{\mathbf{W}}_2$ and $\hat{\mathbf{W}}_3$ clearly target vertical and horizontal differencing of pixels, which smooths out vertical and horizontal edges in the image. The first filter $\hat{\mathbf{W}}_1$ provides further local differencing to smooth out remaining spatial dependencies. The second advantage is that the sparse filters in \eqref{eq:filter} induce a sparse precision matrix on the elicited DGMRF prior, which can accelerate computation. These two advantages can be leveraged for efficient online change detection with image data, as outlined next.

% Figure \ref{fig:deepGMRF} (left) shows, for the above example, the corresponding precision matrix $\mathbf{Q}_{\mathbf{G}}^{-1}$ for the elicited GMRF prior, which can be seen to be highly sparse.

% \begin{figure}[ht]
%     \centering
%     \includegraphics[width = \textwidth]{Q0_construct.png}
%     \caption{Constructing the precision matrix using DGMRF.}
%     \label{fig:Q0_construct}
% \end{figure}

\section{BLAST: Online Change-point Detection}
\label{sec:blast}
Section \ref{sec:prior} provides the image prior $\text{DGMRF}(\hat{\boldsymbol{\Theta}})$ elicited from \textit{offline} reference data. Consider now the \textit{online} setting, where we wish to detect change-points via the sequentially observed data $\mathbf{X}_1, \mathbf{X}_2, \cdots \in \mathbb{R}^{q_1 \times q_2}$. We present in the following an efficient procedure for posterior image updates with online image data, then show how this image posterior can be integrated for efficient computation of the posterior run length distribution, which is used for online change detection. Figure \ref{fig:framework} (right) visualizes this workflow. 

\subsection{Image Posterior Updates}
\label{sec:posterior_update}

We first present a procedure for efficient online posterior image updates. Suppose we observe the sequence of image data $\mathbf{X}_1, \mathbf{X}_2, \cdots$, and further suppose no change-point has occurred (this will be relaxed later). A plausible model on image $\mathbf{X}_t$ is:
\begin{equation}
\text{vec}(\mathbf{X}_t) \overset{i.i.d.}{\sim} \mathcal{N}(\text{vec}(\mathbf{M}), \sigma^2 \mathbf{I}_{p \times p}), \quad t = 1, 2, \cdots,
\label{eq:prechange_model}
\end{equation}
where $\sigma^2 > 0$ is the noise variance for online image observations, and $\mathbf{M}$ is the true underlying pre-change image. Assign to $\mathbf{M}$ the elicited prior $\text{DGMRF}(\hat{\boldsymbol{\Theta}})$ from Section \ref{sec:prior}, or equivalently, its vectorized form $\text{vec}(\mathbf{M}) \sim \mathcal{N}(\boldsymbol{\mu}_0,\mathbf{Q}_0^{-1})$, where the prior mean $\boldsymbol{\mu}_0$ and precision matrix $\mathbf{Q}_0$ follow from \eqref{eq:gmrfexpn} using $\hat{\boldsymbol{\Theta}}$. For scalable change-point detection at each time $t$, we require an efficient online procedure for updating the posterior distribution $\mathbf{M}|\mathbf{X}_1, \cdots, \mathbf{X}_{t-1}$; this is presented below.  

Consider first the posterior $\mathbf{M}|\mathbf{X}_1$ at time $t=1$. It is straight-forward to show that:
\begin{equation}
\mathbf{M}|\mathbf{X}_1 \sim \mathcal{N}(\boldsymbol{\mu}_1,\mathbf{Q}_1^{-1}),
\end{equation}
where its mean vector and precision matrix take the form:
\begin{equation}
     \boldsymbol{\mu}_1 = \mathbf{Q}_1^{-1} \left(\frac{1}{\sigma^2} \text{vec}(\mathbf{X}_1) + \mathbf{Q}_0 \boldsymbol{\mu}_0 \right), \quad 
    \mathbf{Q}_1 = \frac{1}{\sigma^2}\mathbf{I}_{p \times p} + \mathbf{Q}_0.
    \label{eq:update1}
\end{equation}
Note that, once $\mathbf{Q}_1^{-1}$ is computed, the work required to evaluate $\boldsymbol{\mu}_1$ and $\mathbf{Q}_1$ is $\mathcal{O}(p^2)$, where $p$ is the number of image pixels. An efficient procedure for evaluating the matrix inverse $\mathbf{Q}_1^{-1}$ is presented later.

Consider next the posterior $\mathbf{M}|\mathbf{X}_1, \cdots, \mathbf{X}_t$ at time $t = 2, 3, \cdots$. With some manipulations, we can show that its posterior takes the form:
\begin{equation}
\mathbf{M}|\mathbf{X}_1, \cdots, \mathbf{X}_t \sim \mathcal{N}(\boldsymbol{\mu}_t,\mathbf{Q}_t^{-1}),
\label{eq:post}
\end{equation}
where its mean vector and precision matrix can be \textit{recursively} evaluated as:
\begin{equation}
\boldsymbol{\mu}_t = \mathbf{Q}_t^{-1} \left(\frac{1}{\sigma^2} \text{vec}(\mathbf{X}_t) + \mathbf{Q}_{t-1} \boldsymbol{\mu}_{t-1} \right), \quad \mathbf{Q}_{t} = \frac{1}{\sigma^2}\mathbf{I}_{p \times p} + \mathbf{Q}_{t-1}.
\label{eq:update2}
\end{equation}
Here, the mean vector and precision matrix $\boldsymbol{\mu}_{t-1}$ and $\mathbf{Q}_{t-1}$ have already been evaluated at the previous time $t-1$; Equation \eqref{eq:update2} updates such parameters using the current image data. Again, note that once $\mathbf{Q}_t^{-1}$ is computed, the work required to evaluate $\boldsymbol{\mu}_t$ and $\mathbf{Q}_t$ is $\mathcal{O}(p^2)$.

The above, however, relies on an efficient online evaluation of the matrix inverse $\mathbf{Q}_t^{-1}$ at each time $t$. A direct evaluation of such an inverse requires $\mathcal{O}(p^3)$ at \textit{each} time-step, which is clearly unwieldy with large images. The proposition below provides a nice computational shortcut:
\begin{prop} \label{prop1}
    Let $\mathbf{Q}_0 = \mathbf{PDP}^{-1}$ be the eigendecomposition of the precision matrix $\mathbf{Q}_0$. Then we have:
\begin{equation}
\mathbf{Q}_t^{-1} = \mathbf{P}\left(\mathbf{D}+\frac{t}{\sigma^2}\mathbf{{I}}_{p \times p}\right)^{-1}\mathbf{P}^{-1}.
\label{eq:shortcut}
\end{equation}
\end{prop}
\noindent Its proof is provided in Supplementary Materials. The inverse $\mathbf{Q}_t^{-1}$ can thus be efficiently computed as follows. First, perform a one-shot offline Cholesky decomposition of $\mathbf{Q}_0$ to compute its eigendecomposition $\mathbf{Q}_0 = \mathbf{PDP}^{-1}$; more on this below. Next, with this in hand, compute $\mathbf{Q}_t^{-1}$ using Equation \eqref{eq:shortcut}. Note that, since the middle matrix in \eqref{eq:shortcut} is diagonal, such an evaluation requires only $\mathcal{O}(p^2)$ work at each time $t$. Thus, with the Cholesky decomposition of $\mathbf{Q}_0$ performed offline (i.e., before online change detection), $\mathbf{Q}_t^{-1}$ can be computed with $\mathcal{O}(p^2)$ work at each time-step, which reduces the $\mathcal{O}(p^3)$ work for a direct matrix inverse in online implementation.

The remaining step is the offline Cholesky decomposition of the precision matrix $\mathbf{Q}_0$. Note that this only needs to be performed once, prior to online change detection. Recall from Section \ref{sec:dgmrf} that $\mathbf{Q}_0$ is typically highly sparse by the convolutional construction of the deep GMRF. Such sparsity can be exploited to greatly accelerate Cholesky decompositions for large matrices; see \cite{gould2007numerical,scott2023sparse}. These sparse solvers are broadly available in software, e.g., the \texttt{scikit-sparse} library in Python \cite{scikitsparse}, or the \texttt{SparseM} library in \textsf{R} \cite{sparsem}. While the precise complexity of sparse Cholesky decompositions is difficult to pinpoint, empirical experiments have noted a dramatic reduction in runtime \cite{gould2007numerical} over the $\mathcal{O}(p^3)$ cost for dense matrices, particularly as the degree of sparsity increases. Similar speed-ups were observed in our later experiments.

\subsection{Posterior Run Length}

With the updated posterior distribution $\mathbf{M}|\mathbf{X}_1, \cdots, \mathbf{X}_t$ at each time $t$, we then need to use this to compute the desired posterior run length distribution, which guides the online change detection procedure. To do this, we adapt the Bayesian change-point procedure in \cite{BOCD} to develop a new change-point procedure with image data. In what follows, we permit change-points in the sequence of image data $\mathbf{X}_1, \mathbf{X}_2, \cdots$. For the first set of time-steps before the first change-point, we presume its image data follows the sampling model \eqref{eq:prechange_model}, with $\mathbf{M}_{[1]}$ as its underlying mean image. For the next set of time-steps before the second change-point, we similarly presume its image data follows \eqref{eq:prechange_model} with $\mathbf{M}_{[2]}$ as its underlying mean image; this then continues on. We then assign independent priors on the set of underlying images $\mathbf{M}_{[1]}, \mathbf{M}_{[2]}, \cdots \stackrel{i.i.d.}{\sim} \text{DGMRF}(\hat{\boldsymbol{\Theta}})$, to reflect our belief that the elicited image structures are present in the online image data.

With this, let $r_t$ denote the so-called run length at time $t$, defined as the length of time since the last change-point in the sequential procedure. Following \cite{BOCD}, we set the first run time as $r_1 = 0$, then assign the following Markovian prior on the sequence of run times $\{r_t\}_{t \geq 2}$:
\begin{equation}
    [r_t|r_{t-1}] =
    \begin{cases}
      h(r_{t-1} + 1), & \text{if } r_t = 0,\\
      1 - h(r_{t-1} + 1), & \text{if } r_t = r_{t-1}+1, \\
      0, & \text{otherwise}.
    \end{cases}
    \label{eq:rprior}
\end{equation}
Here, $h(\tau)$ is a hazard function that models the probability of a change-point at the current time, given no change-points in the previous $\tau-1$ time-steps. With prior knowledge on how change-points may arise in the system, such knowledge should be carefully used to specify an appropriate hazard function. Without this knowledge, a natural choice may be the constant hazard function $h(\tau) = 1/\lambda$, which corresponds to a memoryless process for change-point arrivals. We use such a memoryless prior in our later experiments, with $\lambda$ set as 20.

% \cmtS{how did we specify $\lambda$?} \cmtX{I set $\lambda = 20$ based on the assumption that our observed time series are short, but I also tried with values up to 100, and the posterior remains largely unchanged. }.

Next, at time $t$, consider the posterior run length distribution $[r_t|\mathbf{X}_{1:t}]$, where $\mathbf{X}_{1:t}$ is shorthand for $\mathbf{X}_{1}, \cdots, \mathbf{X}_t$. This posterior distribution on $r_t$ will be used for online change monitoring in BLAST, as it facilitates probabilistic inference on past change-points. In particular, a concentration of probability at large run lengths suggests no recent change-points, whereas a concentration at small run lengths suggests a recent change-point. For $t=1$, its posterior run length distribution is trivially $r_t|\mathbf{X}_{1:t} \equiv 0$ with probability one. To efficiently compute this posterior distribution for $t \geq 2$, we can rewrite this as:
\begin{equation}
[r_t|\mathbf{X}_{1:t}] = \frac{[r_t,\mathbf{X}_{1:t}]}{[\mathbf{X}_{1:t}]}.
\label{eq:cond}
\end{equation}
Following \cite{BOCD}, the joint distribution $[r_t,\mathbf{X}_{1:t}]$ can then be recursively decomposed as:
\begin{align}
\begin{split}
[r_t, \mathbf{X}_{1:t}] &= \sum_{r_{t-1} \in \mathcal{R}_{t-1}} [r_t,\mathbf{X}_t|r_{t-1},\mathbf{X}_{1:(t-1)}][r_{t-1},\mathbf{X}_{1:(t-1)}]\\
&= \sum_{r_{t-1} \in \mathcal{R}_{t-1}} [\mathbf{X}_t|r_{t-1}, \mathbf{X}_{t,r_t}] [r_t|r_{t-1}] [r_{t-1},\mathbf{X}_{1:(t-1)}]\\
& =: \sum_{r_{t-1}  \in \mathcal{R}_{t-1}} \circled{1} \; \circled{2} \; \circled{3}.
\label{eq:recurse}
\end{split}
\end{align}
Here, $\mathcal{R}_{t-1}$ is the support of $r_{t-1}$, and $\mathbf{X}_{t,r_t}$ denotes the set of $r_t$ past observations at time $t$. Equations \eqref{eq:cond} and \eqref{eq:recurse} form the basis for efficient online updates of the posterior run length distribution, as described next.

Consider the first term $\circled{1}$ in \eqref{eq:recurse}, which concerns the posterior distribution of the current image $\mathbf{X}_t$ conditioned on the past image data $\mathbf{X}_{t,r_t}$ and run length $r_{t-1}$. Suppose $\mathbf{X}_{t,r_t}$ consists of at least one data point, i.e., $r_t \geq 1$; in such a case, all images in $\mathbf{X}_{t,r_t}$ observe (with noise) the same underlying image as $\mathbf{X}_t$. We can thus apply the posterior distribution from Equation \eqref{eq:post} to get:
\begin{equation}
\text{vec}(\mathbf{X}_t)|r_{t-1}, \mathbf{X}_{t,r_t} \sim \mathcal{N}\left(\boldsymbol{\mu}_{t,r_t},\mathbf{Q}_{t,r_t}^{-1} + \sigma^2 \mathbf{I}_{p \times p} \right).
\label{eq:pred}
\end{equation}
Here, $\boldsymbol{\mu}_{t,r_t}$ and $\mathbf{Q}_{t,r_t}$ are the mean vector $\boldsymbol{\mu}_t$ and precision matrix $\mathbf{Q}_t$ using \eqref{eq:post} conditioned on the past image data $\mathbf{X}_{t,r_t}$. Otherwise, if $\mathbf{X}_{t,r_t}$ is empty, i.e., $r_t = 0$, then $\mathbf{X}_t$ is the only observation collected on its underlying image. As such, $\text{vec}(\mathbf{X}_t)|r_{t-1}, \mathbf{X}_{t,r_t}$ simply reduces to its marginal form $\text{vec}(\mathbf{X}_t) \sim \mathcal{N}(\boldsymbol{\mu}_0,\mathbf{Q}_0^{-1}+\sigma^2 \mathbf{I}_{p \times p})$.

Consider next term $\circled{2}$: this immediately follows from the run length prior \eqref{eq:rprior}. Finally, term $\circled{3}$ is the joint distribution of $r_{t-1}$ and $\mathbf{X}_{1:(t-1)}$. As \eqref{eq:recurse} is computed sequentially over time, such a term would have been computed at the previous time-step $t-1$ and can thus be directly plugged in. With these three terms obtained, one can then use Equation \eqref{eq:recurse} to recursively update the joint distribution $[r_t,\mathbf{X}_{1:t}]$ at current time $t$, then Equation \eqref{eq:cond} to compute the desired run length posterior $[r_t|\mathbf{X}_{1:t}]$. Such a recursive procedure provides efficient online updates of the run length posterior to facilitate timely change detection; its full complexity analysis is discussed later.

Online recursive updates of the form \eqref{eq:recurse} fall under the broad class of message-passing algorithms \cite{yedidia2011message}, which are widely used for scaling up cost-intensive algorithms in, e.g., signal processing \citep{donoho2009message} and computer vision \citep{swoboda2017message}. Message-passing algorithms partition a complex quantity of interest (here, the run length posterior $[r_t|\mathbf{X}_{1:t}]$) into simpler terms that are either easy to compute (here, terms $\circled{1}$ and $\circled{2}$) or have already been evaluated at a previous time-step (here, term $\circled{3}$). BLAST leverages this message passing structure of the joint posterior $[r_t,\mathbf{X}_{1:t}]$ over time, together with the efficient recursive updates of the image posterior in \eqref{eq:update2}, to facilitate scalable, probabilistic and structure-aware change detection of large images. In our later experiments with images of dimensions 50 $\times$ 40, the posterior run length can quickly computed in a matter of seconds via this recursive approach. This can be further sped up via GPU computing architecture, which are becoming increasingly available; more on this next.

% The final step of BLAST is to integrate the GMRF prior within the Bayesian change-point detection framework. Suppose the true change-point occurs at time $\tau$. Let \[\mathbf{y}_t^{\text{pre}} \sim \mathcal{N}(\mathbf{x^{\text{pre}}}, \sigma^2\mathbf{\mathrm{I}}_{M}) \quad \text{for} \quad t= 1, \cdots, \tau,\]
% and \[\mathbf{y}_t^{\text{post}} \sim \mathcal{N}(\mathbf{x^{\text{post}}}, \sigma^2\mathbf{\mathrm{I}}_{M}) \quad \text{for} \quad t= \tau+1, \cdots, T.\]
% The goal of detecting the change in the observed images can then be thought of as detecting the differences between the underlying pre-change image $\mathbf{x^{\text{pre}}}$ and the underlying post-change image $\mathbf{x^{\text{post}}}$. We thus desire an approach that investigates the following hypothesis:

% \begin{equation}
% \setlength{\abovedisplayskip}{5pt}
% \setlength{\belowdisplayskip}{5pt}
% H_0 : \mathbf{x^{\text{pre}}} = \mathbf{x^{\text{post}}}, \quad H_A : \mathbf{x^{\text{pre}}} \neq \mathbf{x^{\text{post}}}.
% \label{eq:hypo}
% %\vspace{-0.2in}
% \end{equation}

% To test (\ref{eq:hypo}), we adopt the Bayesian online change-point detection framework in \cite{BOCD}, and present an efficient message-passing algorithm. At each time $t$, within the Bayesian online change-point detection framework, we compute the probability distribution of the time since the last change-point, also known as the run length.

\subsection{Algorithm Statement and Complexity}

\begin{algorithm}[!t]
\caption{BLAST -- Online Image Change-point Detection}\label{alg:alg2}
\begin{algorithmic}
\STATE \textit{\underline{Inputs}}: Online images $\mathbf{X}_1, \mathbf{X}_2, \cdots$ arriving sequentially in time, elicited image prior $\text{DGMRF}(\hat{\boldsymbol{\Theta}})$, hazard function $h(\tau)$. 

\vspace{0.15cm}
\STATE $\bullet$ \hspace{0.2cm} Set the initial run length distribution as $[r_1] \equiv 0$ and $\mathcal{R}_1 \leftarrow \{0\}$. Further initialize $\texttt{var}_{0,0}=1$ and $\texttt{var}_{0,r}=0$ otherwise.\\

$\textbf{for}$ each time-step $t = 1, 2, \cdots$:\\

\STATE \quad $\bullet$ \hspace{0.2cm} Observe image data $\mathbf{X}_t$.\\

\quad $\textbf{for}$ each $r_t \in \mathcal{R}_t$: \hfill \COMMENT{$\mathcal{R}_t := \{0, \cdots, \max \mathcal{R}_{t-1} + 1\}$ if $t > 1$}

% \item $\textbf{for}$ each $r_{t-1}$ in its support set $\mathcal{R}_{t-1}$:

\STATE \quad \quad $\bullet$ \hspace{0.2cm} Initialize $\texttt{var}_{t,r_t} \leftarrow 0$. \hfill \COMMENT{\textit{Running variable to compute $[r_t,\mathbf{X}_{1:t}]$}}

\hspace{0.6cm} $\textbf{for}$ each $r_{t-1} \in \mathcal{R}_{t-1}$:

\STATE \quad \quad \quad \quad $*$ \hspace{0.2cm}  Evaluate $\boldsymbol{\mu}_{t,r_t}$ and $\mathbf{Q}_{t,r_t}$ using the recursive updates from Equation \eqref{eq:update2}.

    \STATE \quad \quad \quad \quad $*$ \hspace{0.2cm} Evaluate $\circled{1} \leftarrow [\mathbf{X}_t|r_{t-1},\mathbf{X}_{t,r_t}]$ using Equation \eqref{eq:pred}.

    \STATE \quad \quad \quad \quad $*$ \hspace{0.2cm} Set $\circled{2} \leftarrow [r_t|r_{t-1}]$ using Equation \eqref{eq:rprior}.
    
    \STATE \quad \quad \quad \quad $*$ \hspace{0.2cm} Retrieve $\circled{3} \leftarrow \texttt{var}_{t-1,r_{t-1}}$ as the previously computed $[r_{t-1},\mathbf{X}_{1:(t-1)}]$.
    
    \STATE \quad \quad \quad \quad $*$ \hspace{0.2cm} Update $\texttt{var}_{t,r_t} \leftarrow \texttt{var}_{t,r_t} + \circled{1} \; \circled{2} \; \circled{3}$.\\
% \vspace{-0.15cm}
\hspace{0.6cm} \textbf{endfor}\\
% \vspace{-0.15cm}
\quad \textbf{endfor}

\STATE \quad $\bullet$ \hspace{0.2cm} Initialize $\texttt{var}_t \leftarrow 0$. \hfill \COMMENT{\textit{Running variable to compute $[\mathbf{X}_{1:t}]$}}\\
\quad $\textbf{for}$ each $r_t \in \mathcal{R}_t$: \\
\quad \quad $\bullet$ \hspace{0.2cm} $\texttt{var}_t \leftarrow \texttt{var}_t + \texttt{var}_{t,r_t}$.\\
\quad \textbf{endfor}\\
\quad $\bullet$ \hspace{0.2cm} Update the posterior run length $[r_t|\mathbf{X}_{1:t}] \leftarrow \texttt{var}_{t,r_t}/\texttt{var}_t$.\\
\textbf{endfor}

\vspace{0.15cm}

% \item Evaluate the posterior mean and variance $\boldsymbol{\mu}_{(t-r_t+1):(t-1)}$ and $\mathbf{Q}_{(t-r_t+1):(t-1)}$ from Equation \eqref{eq:post}.

% \item $ \textbf{Compute}$ growth probabilities:
% \[p(r_t = r_{t-1} + 1, \mathbf{y}_{1:t}) = p(r_{t-1}, \mathbf{y}_{1:t-1}) \pi_t^{(r_{t-1})} (1 - \rm{H}(r_{t-1} + 1)).\]

% \item $ \textbf{Compute}$ change-point probability:
% \[P(r_t = 0, \mathbf{y}_{1:t}) = \sum_{r_{t-1}} p(r_{t-1}, \mathbf{y}_{1:t-1}) \pi^{(r_{t-1})}_t \rm{H}(r_{t-1} + 1).\]

% \item $ \textbf{Compute}$ evidence:
% \[p(\mathbf{y}_{1:t}) = \sum_{r_t} p(r_t, \mathbf{y}_{1:t}).\]

% \item $ \textbf{Compute}$ run length posterior:
% \[p(r_t | \mathbf{y}_{1:t}) = \frac{p(r_t, \mathbf{y}_{1:t})}{p(\mathbf{y}_{1:t})}.\]

% \item $ \textbf{Update}$ model parameters $\mathbf{\mu}_t$ and $\mathbf{Q}_t$:
% \[\mathbf{\mu}_t = \mathbf{Q}_t^{-1} (\frac{1}{\sigma^2}\mathbf{\mathrm{I}} \cdot \mathbf{y}_t + \mathbf{Q}_{t-1}^{-1} \mathbf{\mu}_{t-1})\]
% \[\mathbf{Q}_t = \frac{t}{\sigma^2}\mathbf{\mathrm{I}} + \mathbf{Q}_0.\]

% \item $ \textbf{Update } t \leftarrow t + 1$.
\STATE \textit{\underline{Output}}: The run length posterior $[r_t | \mathbf{X}_{1:t}]$ at times $t=1, 2, \cdots$.
\end{algorithmic}
\end{algorithm}

Algorithm \ref{alg:alg2} summarizes each step of the online change-point detection procedure for BLAST, with Figure \ref{fig:framework} visualizing its workflow. We begin with the elicited image prior from Section \ref{sec:prior}, as well as a choice of hazard function $h(\tau)$ for the change-point prior \eqref{eq:rprior}. The initial run length $r_1$ is set as 0 with probability 1. For each time-step $t$, we first observe the image data $\mathbf{X}_t$, compute the posterior mean vector $\boldsymbol{\mu}_{t,r_t}$ and precision matrix $\mathbf{Q}_{t,r_t}$ via the recursive updates from Equation \eqref{eq:update2}, and evaluate the predictive density $\circled{1}$ using Equation \eqref{eq:pred}. We then evaluate $\circled{2}$ from Equation \eqref{eq:rprior}, then retrieve $\circled{3}$ from the joint density previously computed at time $t-1$. Finally, we compute the joint density $[r_t,\mathbf{X}_{1:t}]$ from Equation \eqref{eq:recurse}, then the run length posterior $[r_t|\mathbf{X}_{1:t}]$ from Equation \eqref{eq:cond}. These steps are iterated at each incoming time-step for online change-point detection.

% We begin with an empty matrix $\rm{R} \in \mathbb{R}^{(T+1) \times (T+1)}$ to store the posterior probabilities of run lengths over time. We then specify the prior mean $\mathbf{\mu}_0$ and prior precision matrix $\mathbf{Q}_0$ on $\mathbf{x^{\text{pre}}}$ as described in Section \ref{prior_elicitation} and the hazard function on change-point. Next, we compute the growth probability and change-point probability utilizing three components: (i) the joint probability message from $t-1$; (ii) the predictive probability $\pi_t^{(r_{t-1})}$; and (iii) the hazard function $\rm{H}(r_{t-1} + 1)$. Finally, the run length probability is computed, and the model parameters are updated as described in Section \ref{posterior_update}. These steps are iterated at each time $t$ to perform online change-point detection. Algorithm \ref{alg:alg2} summarizes the detailed steps of the procedure.

% \subsubsection{Computational Complexity}

We now investigate the computational complexity of BLAST in terms of increasing image dimensionality, i.e., in terms of the number of image pixels $p$. For image prior elicitation (Section \ref{sec:prior}), the primary computational bottleneck lies in the optimization of the ELBO bound \eqref{eq:elbo2}. As + or seq filters are used in the convolutional construction, the evaluation of the log-determinants in \eqref{eq:elbo2} requires only $\mathcal{O}(p)$ work \citep{DGMRF}. Using the stochastic optimization approach described in Section \ref{sec:prior_elicitation}, each iteration of the optimization procedure can then be shown to require $\mathcal{O}(p)$ work. This linear scaling in $p$ permits scalable prior elicitation with large images. While it is difficult to pinpoint the number of optimization iterations needed for convergence (since the ELBO objective is non-convex in $\boldsymbol{\Theta}$ and $\boldsymbol{\Psi}$), in our later experiments, the solver required only several hundred iterations for empirical convergence using the Adam optimizer in \texttt{PyTorch}, which is quite efficient.

Next, for online change-point detection, two steps need to be analyzed. The first is the one-shot Cholesky decomposition of the prior precision matrix $\mathbf{Q}_0$, which is sparse by construction. As discussed in Section \ref{sec:posterior_update}, such sparsity can be exploited to accelerate the usual $\mathcal{O}(p^3)$ cost for decomposing dense matrices. While the theoretical complexity of sparse Cholesky decompositions is difficult to pin down \cite{gould2007numerical}, empirical results (e.g., \cite{scott2023sparse}) show dramatic speed-ups over its dense counterpart. In our experiments, this one-shot Cholesky decomposition required only seconds to perform, and is thus not a major computational bottleneck.

% its precise complexity is difficult to pin down theoretically \cite{gould2007numerical}. We thus denote this runtime as $\texttt{sparseChol}(p)$ in Table \cmtS{ref}, and note that this is much faster than $\mathcal{O}(p^3)$ in practice.

The second step is the computation of the run length posterior using Equation \eqref{eq:recurse} at each time-step $t = 1, 2, \cdots, T$, where $T$ is the total number of considered time-steps. Consider first this computation at a given time-step $t$ and for a fixed run length $r_t > 0$. From \eqref{eq:recurse}, terms $\circled{2}$ and $\circled{3}$ can be evaluated with $\mathcal{O}(1)$ work. Using the recursive approach in Equation \eqref{eq:update2} along with Proposition \ref{prop1}, term $\circled{1}$ can be computed in $\mathcal{O}(p^2)$ work. Next, combining these terms in \eqref{eq:recurse} takes $\mathcal{O}(1)$ work, since there can only be one choice of $r_{t-1}$ leading to $r_t > 0$. Thus, for \textit{each} $r_t > 0$, the computation of its joint distribution $[r_t,\mathbf{X}_{1:t}]$ requires $\mathcal{O}(p^2)$ work, and hence its computation for \textit{all} $r_t>0$ requires $\mathcal{O}(Tp^2)$. A similar argument for $r_t=0$ shows that the computation of its joint distribution $[r_t,\mathbf{X}_{1:t}]$ requires $\mathcal{O}(Tp^2)$ work as well. Finally, putting everything together, the computation of the full run length posterior at each time $t$ requires $\mathcal{O}(Tp^2)$ work. Given a fixed number of time-steps $T$, this runtime can be expressed as $\mathcal{O}(p^2)$. This quadratic runtime on $p$ for posterior run length updates facilitates \textit{scalable} online change detection with image data. 

In the case of many considered time-steps $T$ (e.g., the monitoring of a manufacturing process for thousands of time-steps), one can further reduce computation of the posterior run length by imposing a reasonable upper bound $R \ll L$ on the run length distribution; this bypasses the need for computing the posterior for run lengths $r_t > R$. With this, the posterior run length can then be updated at each time with $\mathcal{O}(Rp^2)$ work. Such a ``window-limited'' approach has been used for accelerating recent change detection methods \cite{zheng2023percept,xie2021sequential}.

% We thus summarize this runtime as $\mathcal{O}(p^2)$ for each time $t$ in Table \cmtS{ref}.

% We investigate next the computational complexity of BLAST, in terms of images with $M$ pixels as well as time series length $T$. For the filter optimization (Algorithm \ref{alg:alg1}), it requires $\mathcal{O}(M)$ work for a fixed number of iterations of optimization \cite{DGMRF}. The covariance matrix construction requires again $\mathcal{O}(M)$ work to map through all pixels in the images. For the Bayesian change-point (Algorithm \ref{alg:alg2}), the posterior computation at each time $t$ requires $\mathcal{O}(M^3)$ work without applying Proposition \ref{prop1}. When applying Proposition \ref{prop1} for online updating of image posterior, it requires $\mathcal{O}(M^2)$ work at each time $t$, and a one-shot computation of the eigendecomposition of $\mathbf{Q}_0^{-1}$ that requires $\mathcal{O}(M^3)$ work. Summarizing the above, the Bayesian change-point (Algorithm \ref{alg:alg2}) requires an initial $\mathcal{O}(M^3)$ work for eigendecomposition, and $\mathcal{O}(T^2M^2)$ for posterior computation for the entire time series. Clearly, this is not optimal for a time series with a large $T$, and a simple solution is to place a cap on the possible values of $r_t$, which reduces the computational complexity to $\mathcal{O}(TM^2)$.

\section{Simulation Experiments}
\label{sec:num}
We now compare BLAST with existing benchmarks in a suite of simulation experiments. These experiments explore a variety of image change-point scenarios, including spatial correlation, edge structure and pixel intensity changes.

% Here, the underlying images are generated with edge-structure. We investigate several change scenarios, including background correlation change, structure change and intensity change. Here, the simulation of the true images $\mathbf{X}$ follows an appropriately-defined Gaussian Markov Random Field (GMRF) model; more details will be provided later for each scenario. 

\begin{figure}[!t]
    \centering
    \includegraphics[width = 0.8\textwidth]{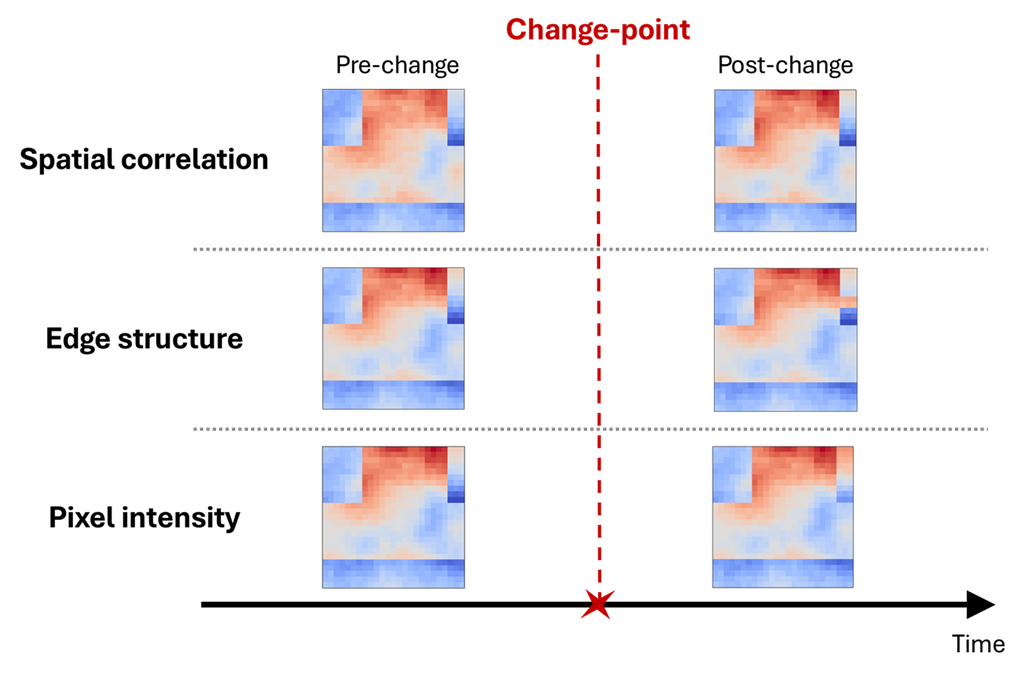}
    \caption{Visualizing the pre-change and post-change true images for the three considered changes (on spatial correlation, pixel intensity and edge structure) in our simulation experiments. Here, the true change point $t^* = 50$ is marked by the red `x'.}
    \label{fig:sim_change}
\end{figure}

The general simulation set-up is as follows. For all experiments, we consider a total of $T=100$ time-steps, with a single true change-point at time $t^* = 50$. Here, the true underlying pre-change and post-change images (denoted $\mathbf{M}_{\rm pre}$ and $\mathbf{M}_{\rm post}$, respectively; both with $25 \times 25$ pixels) have well-defined image structures in terms of edges and shapes (see Figure \ref{fig:sim_change}). The noisy sequence of image data $\mathbf{X}_1, \cdots, \mathbf{X}_T$ is then generated with i.i.d. $\mathcal{N}(0,1)$ noise on each image pixel. The goal is to investigate how well each method leverages this underlying image structure for quick and probabilistic online detection of the change at time $t^* = 50$.

The following benchmarks introduced in Section \ref{sec:mot} are used here for comparison:
\begin{itemize}
    \item \textbf{Hotelling-$T^2$} \cite{hotelling1947multivariate}: Here, the Hotelling-$T^2$ is implemented as described in Section \ref{sec:existing_methods}. We first apply the PCA approach in \cite{xie2021sequential} to extract its top 15 principal components. The online image data are then projected by these principal components to yield scores to use within the monitoring statistic \eqref{eq:hotelling}. The drift constant $d^{\rm H}$ is estimated via the same approach in \cite{zheng2023percept,xie2021sequential}. Details can be found in Section \ref{sec:mot}.
%     The second is the parametric Hotelling's $T^2$ test, using the $15$ extracted principal components from PCA on the original data. 
    \item \textbf{MMD} \citep{mmd}: The MMD approach is similarly implemented as described in Section \ref{sec:mot}. Here, we adopt the standard isotropic Gaussian kernel for $k$ within the MMD, as suggested in \cite{zheng2023percept}. The kernel lengthscale parameter for $k$ is then fitted via the so-called ``median trick'' \citep{med_trick}, i.e., set as the median of the pairwise distances between data points. Details on this can be found in Section \ref{sec:mot}.
    % The last method is the maximum mean discrepancy (MMD) test. In our implementation, we used the standard Gaussian radius basis function (RBF) kernel $K(\cdot,\cdot)$, where the kernel bandwidth is chosen using the so-called ``median trick'' \citep{med_trick}, i.e., set to be the median of the pairwise distances between data points.
    \item \textbf{MONAD} \cite{MONAD_Doshi2021}: Recall that MONAD is a recent deep-learning-based approach for image change detection. MONAD relies on a generative adversarial network for frame prediction, along with a deep-learning object detector for feature extraction. Our experiments make use of the source code provided by the authors in \cite{MONAD_Doshi2021}.
    % The third is the Multi-Objective Neural Anomaly Detector (MONAD) proposed in \cite{MONAD_Doshi2021}, which makes use of the deep learning-based features to construct the test statistics.
        
    % With this, un-BLAST follows the same steps as BLAST .
    \item \textbf{BLAST}: This is the full BLAST procedure in Sections \ref{sec:prior} and \ref{sec:blast}. We first perform image prior elicitation following Algorithm \ref{alg:alg1}, with $L=3$ layers and $3 \times 3$ seq filters for the DGMRF. Here, the offline reference dataset used for elicitation consists of 30 randomly generated images, independently simulated from the pre-change model. We then perform the online change detection procedure in Algorithm \ref{alg:alg2}. 
    % Code for our full implementation is provided in Supplementary Materials.

    \item \textbf{un-BLAST}: This is a baseline that investigates the importance of the prior elicitation procedure from Section \ref{sec:prior}. un-BLAST uses the same BLAST procedure from Section \ref{sec:blast}, but with the simple ``unstructured'' prior on the underlying images $\text{vec}(\mathbf{M}_{[1]}), \text{vec}(\mathbf{M}_{[2]}), \cdots \stackrel{i.i.d.}{\sim} \mathcal{N}(\tilde{\boldsymbol{\mu}},\tilde{\mathbf{D}})$, $\tilde{\mathbf{D}} = \text{diag}\{\tilde{\mathbf{d}}\}$, where $\tilde{\boldsymbol{\mu}}$ and $\tilde{\mathbf{d}}$ are elicited from historical reference data via marginal likelihood maximization. Such a prior is unstructured in the sense that it does not capture image structure, whereas the DGMRF prior (used in BLAST) does. This baseline can thus shed light on how an elicited image-structure-aware prior can facilitate quicker online change detection.
\end{itemize}

\subsection{Spatial Correlation Change}
\label{sec:spatchange}

Consider first the set-up for the spatial correlation change experiment. From Figure \ref{fig:sim_change} (top), we see that both the pre-change and post-change images have clearly defined edge structures in terms of its three rectangular panels. The pre-change image is generated as follows. Within the three panels, we assign small pixel intensity values (blue in Figure \ref{fig:sim_change}); outside of such panels, we assign large pixel intensities (red in Figure \ref{fig:sim_change}). We then add on a spatially correlated layer $\mathbf{Z} \in \mathbb{R}^{q_1 \times q_2}$, sampled from a zero-mean Mat\'ern GMRF \citep{lindgren2011explicit} model. More specifically, $\mathbf{Z}$ is sampled using the precision matrix $\zeta^2 ((\kappa^2\mathbf{I} + \mathbf{G})^\gamma)^\intercal
((\kappa^2\mathbf{I} + \mathbf{G})^\gamma)$, with $\gamma=1$, $\zeta=1$ and $\kappa = \sqrt{8}/20$. Here, $\kappa$ is an inverse length-scale parameter that controls the degree of spatial correlation in $\mathbf{Z}$. The post-change image is generated in a similar fashion with the same three blocks, but its spatially correlated layer is simulated with a smaller $\kappa=\sqrt{8}/30$. This induces a subtly greater degree of spatial correlation compared to the pre-change image (see Figure \ref{fig:sim_change} top). The noisy pre-change and post-change image observations are then sampled with i.i.d. $\mathcal{N}(0,1)$ noise, with the true change-point at time-step $t^*=50$.

% We first consider the case of spatial correlation changes, where the pre-change and post-change data are sampled with noise $\mathcal{N}(0, 1)$ from the $25 \times 25$ true images shown in Figure \ref{fig:correlation_data}. Specifically, the spatial correlation distance $l$ of the underlying GMRF is set to $20$ for the pre-change images and $30$ for the post-change images, where a larger correlation distance $l$ results in greater smoothness in the images. To introduce specific visual structures, predefined edge patterns are overlaid on both the pre- \and post-change GMRF images (Figure \ref{fig:correlation_data}).

\begin{figure}[!t]
   \centering
   \includegraphics[width=\textwidth]{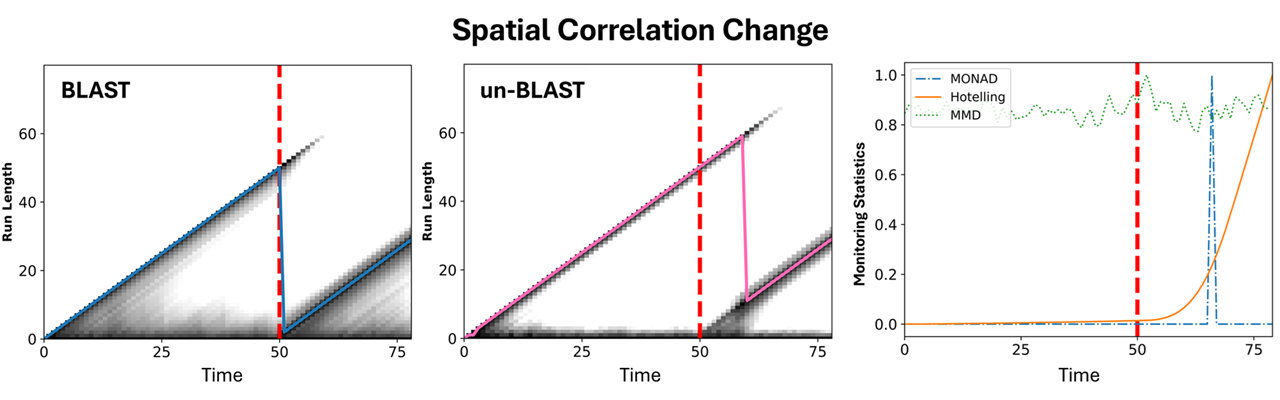}
% \begin{tabular}{ccc}
%     \includegraphics[width=0.33\textwidth]{edges_sim/correlation_change/BLAST_results.png}
%     \includegraphics[width=0.33\textwidth]{edges_sim/correlation_change/bayes_NIP_results.png}
%     \includegraphics[width=0.33\textwidth]{edges_sim/correlation_change/benchmark_results_downsized.png}\\
%     (a) \hspace{1.5in} (b) \hspace{1.5in} (c)
%     \end{tabular}
    \caption{
    Monitoring statistics of the compared methods for the spatial correlation change experiment, with the true change-point at time $t^*=50$ dotted in red. (Left) and (Middle) show the run length posterior for BLAST and un-BLAST, with its maximum a posterior run length $r_t^{\rm MAP}$ marked by solid lines. (Right) shows the monitoring statistics from existing methods.
    % Results for the setting of spatial correlation change, and the vertical red dashed line indicates the true change-point: (a) The run length posterior at each time $t$ for BLAST using a logarithmic color scale. Darker indicates higher probability. The saw-toothed lines are the maximum probability at each time $t$ for BLAST. (b). Same for BLAST-unstr. (c) Test statistics using MONAD, Hotelling's $T^2$ and MMD at each time $t$.
    }
    \label{fig:correlation_results}
\end{figure}

Figures \ref{fig:correlation_results} (left) and (middle) show the run length posterior distribution for BLAST and un-BLAST, with its maximum a posteriori (MAP) run length $r_t^{\rm MAP}$ marked by solid lines. For BLAST, we see that its run length posterior increases steadily with high probability (along with its MAP) before the change-point, which is as desired. After the change-point at $t^*=50$, its run length posterior quickly dips with high probability (along with its MAP) to zero, which shows the proposed method can indeed quickly identify this change-point. For un-BLAST, its run length posterior also increases steadily with high probability (along with its MAP) before the change-point as desired. However, after the change-point at $t^*=50$, its run length posterior (along with its MAP) experiences considerable delay before dipping to zero, thus indicating delayed change detection. This shows that, with a careful prior elicitation of desired image structure, BLAST can leverage such structure to quickly identify abrupt changes in image data.

% This is noticeably different for un-BLAST, which uses the same online framework as BLAST but without the elicited image prior from Section \ref{sec:prior}. Here, after the change-point, we see its run length posterior gradually dips, but with considerably greater delay than BLAST. This shows that, 

Figure \ref{fig:correlation_results} (right) shows the corresponding detection statistics from existing methods. Here, the Hotelling-$T^2$ again experiences considerable detection delay: its monitoring statistic increases slowly after the change-point at $t^*=50$. MMD similarly has difficulties in identifying the desired change-point. This is not too surprising, as both methods largely ignore the underlying image structure, which may lead to delayed detection (Challenge \iilink). For MONAD, we observe not only delayed detection but also noticeable instabilities in its monitoring statistics after the change-point. As mentioned in Section \ref{sec:mot}, this may be due to the lack of uncertainty quantification on the learned image structure (Challenge \iiilink), which results in inflated model confidence and thus unstable monitoring. Compared to these existing methods, BLAST offers quicker detection of the underlying change, along with a quantification of monitoring uncertainty via its posterior run length.

% \begin{figure}
%     \centering
%     \includegraphics[width=0.5\linewidth]{edges_sim/correlation_change/ARL-EDD.png}
%     \caption{Caption}
%     \label{fig:enter-label}
% \end{figure}

% For BLAST, we observe a consistent increment by $1$ in the run length with the maximum probability before $t=50$, suggesting that no change occurs. The run length with the maximum probability then suddenly drops back to $0$ at $t=51$, indicating an immediate detection without delay. In contrast, while BLAST-unstr does manage to detect this change, it suffers from a significant detection delay indicated by the maximum probability. The abrupt reduction of the run length to $0$ at $t=51$ signals a change-point. However, the probability of no change continuing to dominate until roughly $t=60$, suggests the necessity of integrating the common image structures in the prior precision matrix for faster and more accurate detection of change points. Figure \ref{fig:correlation_results}(c) shows the test statistics from MONAD, Hotelling's $T^2$ and MMD. The MONAD statistics is not able to detect the change at all because it does not consider image feature uncertainty. In addition, there are no discernible objects in the data, and thus the pre-trained object detector fails. The Hotelling's $T^2$ statistic increases noticeably more slowly after the change, which suggests a larger detection delay. For the MMD statistic, we see that while it peaks up after the change-point, its pre-change statistic is quite unstable and volatile, which leads to increased false alarms. 

\subsection{Edge Structure Change}
\label{sec:edgechange}

Consider next the set-up for the edge change experiment. Here, the pre-change image here is the same as the pre-change image from Section \ref{sec:spatchange}. Instead of a change in spatial correlation, however, the post-change image here features a split of the top-right block into two separate blocks (see Figure \ref{fig:sim_change} middle), resulting in a change of edge structure. The same spatial correlation layer is used for both the pre-change and post-change image. As before, the noisy pre-change and post-change image observations are sampled with i.i.d. $\mathcal{N}(0,1)$ noise, with the true change-point occurring at $t^* = 50$.

% Figure \ref{fig:correlation_data}(b). Here, the true pre-change image with $25 \times 25$ pixels is generated as before, with the spatial correlation layer simulated using $\zeta=50$. The post-change image is then simulated with the same spatial correlation layer, but with the top-right block split into two blocks (see Figure \ref{fig:correlation_data}(b)). The true change-point thus involves a change in the underlying image structure (here, image blocks). From this, the noisy pre-change and post-change image data are again sampled with i.i.d. $\mathcal{N}(0,1)$ noise, with the true change-point happening at $t^* = 50$.

% , the case of the visual structure change, where the pre- and post-change data are sampled with noise $\mathcal{N}(0, 1)$ from the $25 \times 25$ true images shown in Figure \ref{fig:structure_data}. The true images are generated using GMRF with the spatial correlation distance $l = 50$. The predefined edge patterns are overlaid on the GMRF pre- \and post-change images, with a segment missing from the upper-right block edge for post-change, as seen in Figure \ref{fig:structure_data}. 

% \begin{figure}[h]
%     \centering
%     \includegraphics[width = 0.8\textwidth]{edges_sim/structure_change/structure_samples.png}
%     \caption{Visualizing the pre-change and post-change true images along with their generated samples: complete upper-right block edge (left) and a segment missing from the upper-right block edge (right).\cmtS{integrate into one figure, see above comment}}
%     \label{fig:structure_data}
% \end{figure}

\begin{figure}
    \centering
% \begin{tabular}{ccc}
%     \includegraphics[width=0.33\textwidth]{edges_sim/structure_change/BLAST_results.png}
%     \includegraphics[width=0.33\textwidth]{edges_sim/structure_change/bayes_NIP_results.png}
%     \includegraphics[width=0.33\textwidth]{edges_sim/structure_change/benchmark_results_downsized.png}
%     \\
%     (a) \hspace{1.5in} (b) \hspace{1.5in} (c)
%     \end{tabular}
\includegraphics[width=\textwidth]{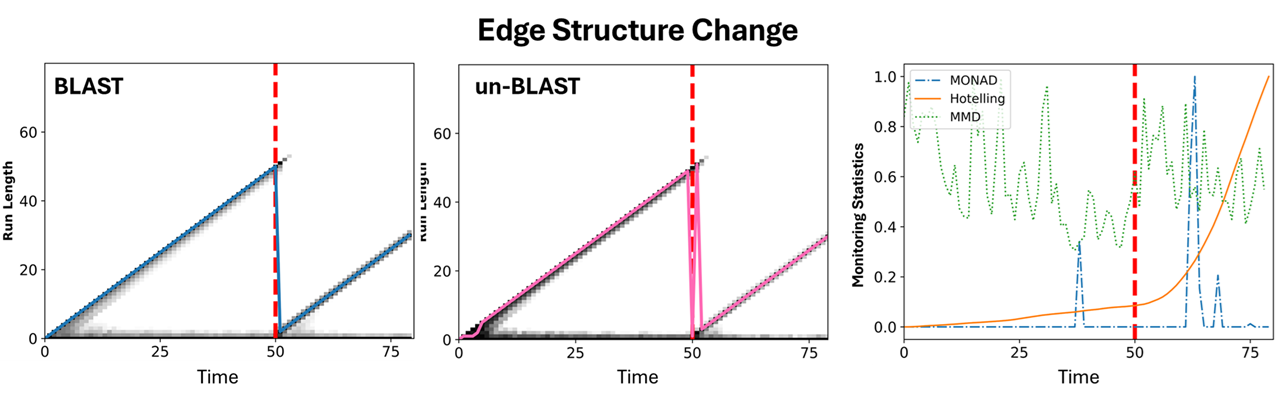}
    \caption{
    Monitoring statistics of the compared methods for the edge structure change experiment, with the true change-point $t^*=50$ dotted in red. (Left) and (Middle) show the run length posterior for BLAST and un-BLAST, with its maximum a posterior run length $r_t^{\rm MAP}$ marked by solid lines. (Right) shows the monitoring statistics from existing methods.
    % Results for the setting of structure change, and the vertical red dashed line indicates the true change-point: (a) The Run length posterior at each time $t$ for BLAST using a logarithmic color scale. Darker indicates higher probability. The saw-toothed lines are the maximum probability at each time $t$ for BLAST. (b). Same for BLAST-unstr. (c) Test statistics using MONAD, Hotelling's $T^2$ and MMD at each time $t$.
    }
    \label{fig:structure_results}
\end{figure}

Figure \ref{fig:structure_results} (left) and (middle) show the run length posterior for BLAST and un-BLAST, with its maximum a posteriori run length $r_t^{\rm MAP}$ marked by solid lines. For BLAST, its run length posterior (along with its MAP) again increases steadily before the change-point and dips quickly to zero after, which suggests it can quickly detect this edge structure change. For un-BLAST, we see a slight delay in identifying this change-point, which again points to the importance of a carefully elicited image-structure-aware prior. Figure \ref{fig:structure_results} (right) shows the monitoring statistics from existing methods. Similar observations hold from before. Both the Hotelling-$T^2$ and MMD have difficulties in quickly identifying the change-point, as such methods largely ignore the underlying image structure. MONAD again shows considerable instabilities: its monitoring statistics not only peak up several times well after the change-point, but also before the change-point. This can be attributed to its lack of UQ in image learning (Challenge \iiilink), which results in model overconfidence and unstable monitoring performance.

% Figure \cmtS{add} affirms these findings via its ARL-EDD plots. \cmtS{a few sentences describing}

% \begin{figure}
%     \centering
%     \includegraphics[width=0.5\linewidth]{edges_sim/structure_change/ARL-EDD.png}
%     \caption{Caption}
%     \label{fig:enter-label}
% \end{figure}

% We see that the run length with the maximum probability for BLAST has a consistent increase by $1$ before $t=50$, suggesting that no change occurs. This run length sharply resets to $0$ at $t=51$ indicating an immediate detection of change without delay. A similar pattern is observed for BLAST-unstr, though the run length which shows the highest probability exhibits some instability in the initial time points. Around $t=50$, there is noticeable fluctuation in the belief between the presence and absence of a change-point. Figure \ref{fig:structure_results}(b) shows the test statistics from MONAD, Hotelling's $T^2$ and MMD. The Hotelling's $T^2$ statistics shows a relatively larger detection delay. Neither MONAD nor MMD statistics succeeds in detecting the change.

%By adopting the Deep GMRF framework, we trained a CNN by minimizing a combination of data fidelity and regularization terms to get the optimal filters (Figure \ref{fig:filters}). As seen in Figure \ref{fig:cnn}, the first filter learns the vertical edges from the data, the second filter is close to an identity function, and the third filter learns the horizontal edges. The precision matrix stencil is then computed using the optimal filters, and the precision matrix can then be computed. We take the resulting precision matrix as the prior precision matrix for cdhange-point detection. 

\subsection{Intensity Change}

Consider finally the set-up for the pixel intensity change experiment. Again, the pre-change image here is the same as the pre-change image from Section \ref{sec:spatchange}. The post-change image here features an increase in pixel intensity within the top-right block. The same spatial correlation layer is again used for both the pre-change and post-change image. The noisy pre-change and post-change image observations are sampled with i.i.d. $\mathcal{N}(0,1)$ noise, with the true change-point occurring at $t^* = 50$.

% in Figure \ref{fig:correlation_data}(c). Here, the true pre-change $25 \times 25$ image is generated as before, with the spatial correlation layer simulated with $\zeta = 50$. The post-change image is then simulated with the same spatial correlation layer, but with the pixel intensities within the top-right block increased. The true change-point thus involves a local change in the underlying image intensity. From this, noisy pre-change and post-change image data are then sampled with i.i.d. $\mathcal{N}(0,1)$ noise, with the true change-point at $t^*=50$.

\begin{figure}[!t]
    \centering
    \includegraphics[width=\textwidth]{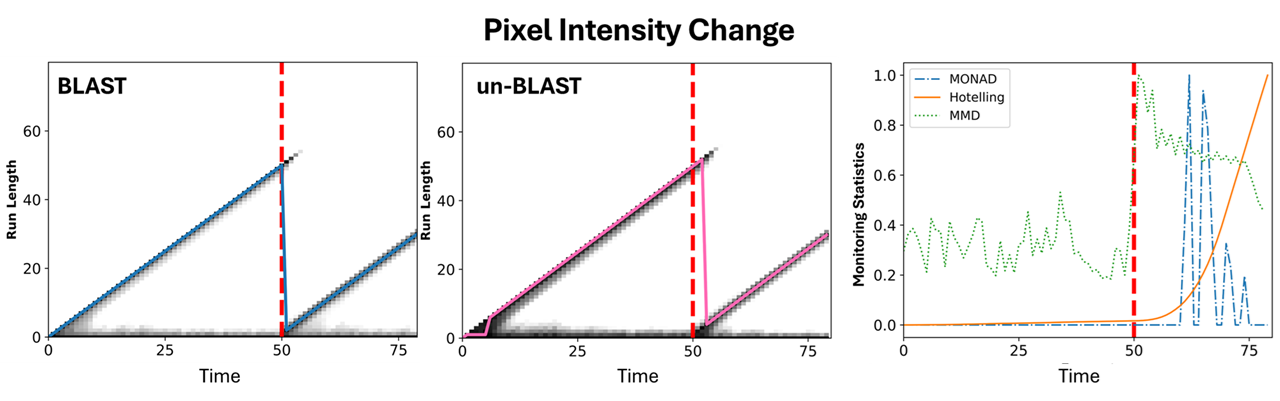}
% \begin{tabular}{ccc}
%     \includegraphics[width=0.33\textwidth]{edges_sim/intensity_change/BLAST_results.png}
%     \includegraphics[width=0.33\textwidth]{edges_sim/intensity_change/bayes_NIP_results.png}
%     \includegraphics[width=0.33\textwidth]{edges_sim/intensity_change/benchmark_results_downsized.png}
%     \\
%     (a) \hspace{1.5in} (b) \hspace{1.5in} (c)
%     \end{tabular}
    \caption{Monitoring statistics of the compared methods for the intensity change experiment, with the true change-point $t^*=50$ dotted in red. (Left) and (Middle) show the run length posterior for BLAST and un-BLAST, with its maximum a posterior run length $r_t^{\rm MAP}$ marked by solid lines. (Right) shows the monitoring statistics from existing methods.}
    \label{fig:intensity_results}
\end{figure}

Figure \ref{fig:intensity_results} (left) and (middle) show the run length posterior for BLAST and un-BLAST. As before, we see that BLAST quickly detects this change: its run length posterior (along with its MAP) increases steadily before and dips quickly after the change-point. un-BLAST similarly experiences a slight delay in identifying this change, which points to the importance of integrating elicited image structure as prior information. Figure \ref{fig:intensity_results} (right) shows the monitoring statistics from existing methods. Here, both the Hotelling-$T^2$ and MONAD are considerably delayed in detecting the change, with the latter also experiencing considerable instabilities. Unlike previous experiments, the MMD provides quick detection of this change-point: its monitoring statistics spike up quickly after the change. Such a method, however, does not yield the desired quantification of monitoring uncertainty provided by BLAST via its posterior run length.

\section{Applications} 
We now investigate the performance of BLAST in two practical applications. The first is the earlier street scene monitoring application from Section \ref{sec:mot}, and the second involves online process monitoring for metal additive manufacturing.

\subsection{Street Scene Monitoring}
\label{sec:street}

Consider first the street scene monitoring application from Section \ref{sec:mot}. Our set-up is adopted from the data collected in \cite{car_data}. There, image snapshots are taken from a static USB camera, looking down on a two-lane street with bike lanes and pedestrian sidewalks (see Figure \ref{fig:motivation}). These snapshots contain highly structured image features in the background, e.g., trees, streets and road markings. This presents a challenging case study for BLAST, in testing whether it can leverage such image structure for quick change detection. We investigate in particular a sequence of $T=35$ snapshots, with the considered change-point taking the form of a car entering the scene at time $t^* = 16$. Here, the same methods are compared as in earlier experiments. For BLAST, the image prior (see Section \ref{sec:prior}) is elicited from a separate reference set of $N=30$ image snapshots taken from \cite{car_data}, which contained similar infrastructure features. All methods follow the same implementation as described in Section \ref{sec:num}.

% See Figure \ref{fig:car_data} for snapshots with a car driving starting from $t^* = 16$. 

 % the infrastructure features, such as bike lanes and pedestrian sidewalks are captured using $30$ training images of the static street scenes with edge detector, and are then integrated in the prior precision matrix for quick detection

% The Hotelling's $T^2$ test is applied using the $15$ extracted principal components from PCA on the image data, and the MONAD statistics is performed on the extracted deep-learning features. The MMD test is performed on the image data directly, with the RBF kernel, and the bandwidth is selected using the ``median trick" as described in Section \ref{sec:num}. 

\begin{figure}[!t]
    \centering
    \includegraphics[width=\textwidth]{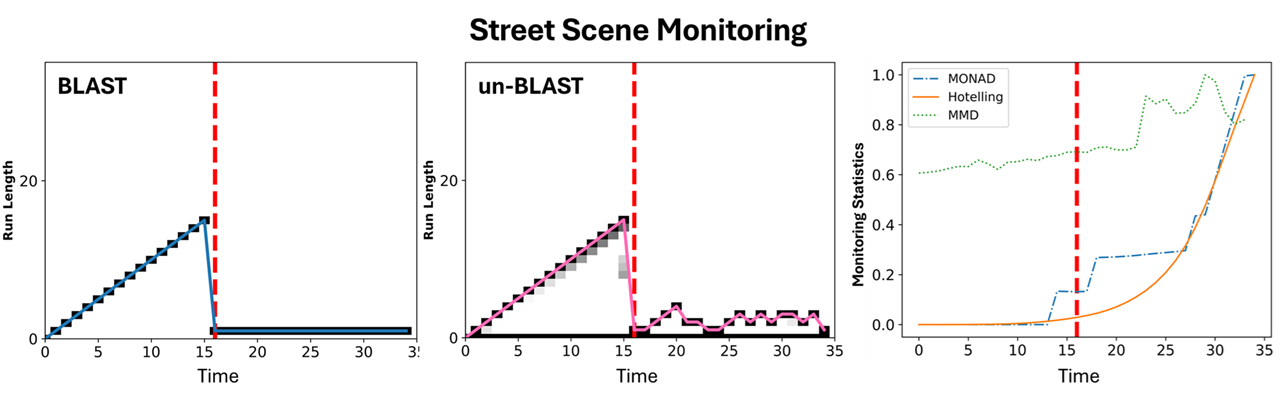}
% \begin{tabular}{ccc}
%     \includegraphics[width=0.33\textwidth]{Car/car_BLAST.png}
%     \includegraphics[width=0.33\textwidth]{Car/car_NIP.png}
%     \includegraphics[width=0.33\textwidth]{Car/car_benchmark.png}
%     \\
%     (a) \hspace{1.5in} (b) \hspace{1.5in} (c)
%     \end{tabular}
    \caption{Monitoring statistics of the compared methods for the street scene monitoring application, with the true change-point $t^*=16$ dotted in red. (Left) and (Middle) show the run length posterior for BLAST and un-BLAST, with its maximum a posterior run length $r_t^{\rm MAP}$ marked by solid lines. (Right) shows the monitoring statistics from existing methods.}
    \label{Car_results}
\end{figure}

Figures \ref{Car_results} (left) and (middle) show the run length posteriors for BLAST and un-BLAST. We see that the run length posterior for BLAST increases steadily with high probability (along with its MAP) prior to the change-point, as desired. It then dips down to zero with high probability (along with its MAP) promptly after the change-point, thus providing quick change detection. For un-BLAST, which does not leverage the elicited image structure as prior information, its run length posterior similarly increases steadily prior to the change-point. However, after the change (i.e., as the car arrives and passes through the scene), its run length posterior fluctuates considerably from zero with high probability. This is rather undesirable: as the car continues to pass through after time $t^*=16$, the scene continues to change and thus its run length should be close to zero with high probability for \textit{all} times after $t^*$. One plausible reason is that, without prior knowledge on expected image structure, it may be difficult for un-BLAST to quickly detect iterative changes. To contrast, with a careful prior elicitation of image structure, BLAST achieves this desired outcome of keeping the run length posterior near zero with high probability at all times after $t^*$.

Figure \ref{Car_results} (right) shows the monitoring statistics for existing methods. Recall our analysis of this from Section \ref{sec:mot}, where we observed several limitations: the Hotelling-$T^2$ and MMD experience delayed detection performance as they largely ignore underlying image structure, whereas MONAD yields unstable monitoring statistics due to its lack of uncertainty quantification on image structure. BLAST addresses both challenges via a careful elicitation and integration of the image prior model within the Bayesian change-point procedure. In doing so, Figure \ref{Car_results} shows that BLAST enjoys quicker detection performance with greater stability over existing methods, along with a probabilistic quantification of monitoring uncertainty for informed decision-making.

\subsection{Process Monitoring for Metal Additive Manufacturing}

Consider next a process monitoring application for metal additive manufacturing. Additive manufacturing plays a fundamental role in the so-called ``Industry 4.0'' \citep{lasi2014industry}, the fourth industrial revolution that marks rapid advancements of manufacturing technologies in the 21st century. In contrast with traditional manufacturing, which subtracts or forms material into the shape of a product, additive manufacturing \citep{gibson2021additive} instead makes use of the layer-by-layer fusing, melting or bonding of material (e.g., powders or plastics) for three-dimensional product construction. A key advantage is that such manufacturing is directly guided by computer-aided design software and requires little human intervention. Metal additive manufacturing \citep{armstrong2022overview}, in particular, has had a remarkable impact in advancing broad industries, including surgical \citep{chen2021function,chen2022adaptive}, aerospace \citep{gradl2021process} and aviation \citep{blakey2021metal} applications. 

\begin{figure}[!t]
    \centering
    \includegraphics[width=0.9\linewidth]{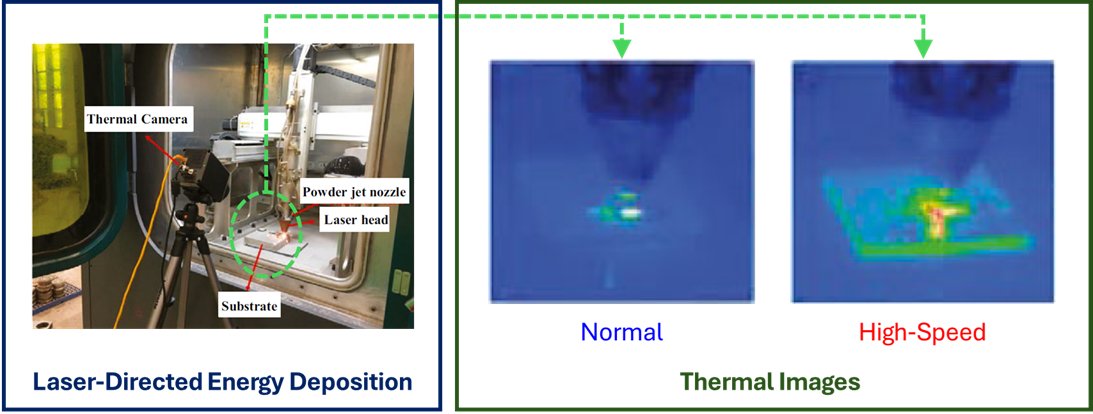}
    \caption{(Left) The laser-directed energy deposition process in \cite{zhang2022review}, with experimental equipment for thermal imaging. (Right) Thermal images of the process under normal operating conditions and at higher-than-expected laser beam speed. Figures adapted from Figure 17 of \cite{zhang2021numerical}.}
    \label{fig:lded}
\end{figure}

We investigate next the real-time monitoring of a laser-directed energy deposition process (LDED; \citep{lia2018thermal}), a common metal additive manufacturing method that blows powder into a melt pool formed by a moving laser beam to produce a layer-by-layer construction of the product. Figure \ref{fig:lded} (left) shows the LDED manufacturing process (taken from \cite{zhang2022review}), along with experimental equipment for real-time monitoring via thermal imaging. Here, a careful control of process parameters, e.g., laser beam speed, is critical for ensuring product quality \citep{zhang2021numerical}: an anomalously high laser speed may cause poor wettability of the molten liquid flow and weak interactions between laser beam and powder particles, which results in deteriorated process efficiency. Figure \ref{fig:lded} (right) shows the thermal images taken under normal operating conditions and at higher-than-expected laser beam speed. We see that such an anomaly results in clear visual changes in the thermal images, and the goal is to quickly detect such changes to ensure product quality. 

This case study nicely captures the three motivating challenges outlined in Section \ref{sec:intro}. First, with high-dimensional and high-frame-rate imaging systems, monitoring statistics need to be computed in a scalable fashion to avoid delays in change detection (Challenge \ilink). Second, as shown in Figure \ref{fig:lded} (right), the monitored images exhibit distinct image structure, e.g., concentrated pixels surrounded by diffuse blotches, that should be leveraged for quick monitoring of potential anomalies (Challenge \iilink). Finally, a probabilistic quantification of monitoring uncertainty (Challenge \iiilink) is essential here. Since interruptions in a manufacturing process (e.g., for fault maintenance) are considerably costly, such a quantification provides engineers with a measure of certainty that a change has indeed occurred prior to interrupting the process.

\begin{figure}[!t]
    \centering
    \includegraphics[width=\textwidth]{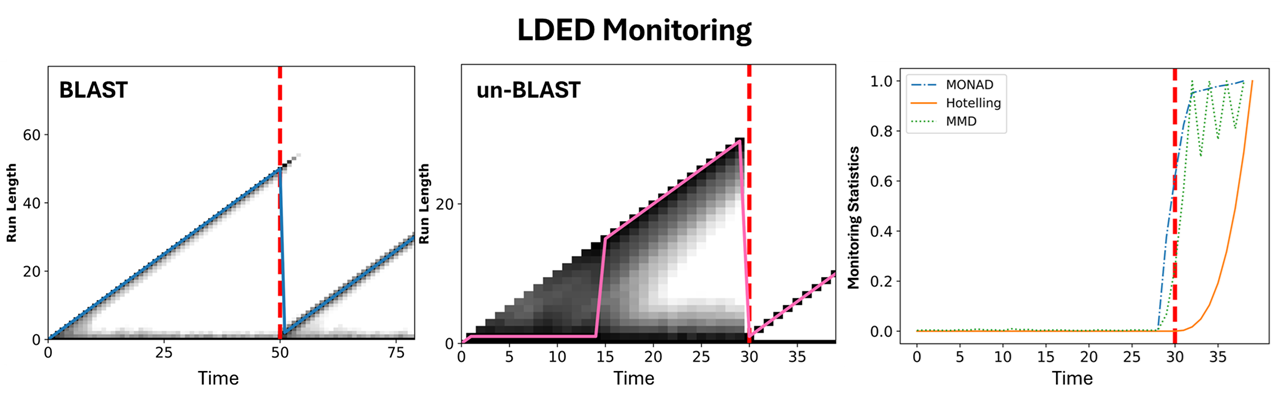}
% \begin{tabular}{ccc}
%     \includegraphics[width=0.33\textwidth]{LaserEtching/LaserEtching_BLAST.png}
%     \includegraphics[width=0.33\textwidth]{LaserEtching/LaserEtching_NIP.png}
%     \includegraphics[width=0.33\textwidth]{LaserEtching/LaserEtching_benchmark.png}
%     \\
%     (a) \hspace{1.5in} (b) \hspace{1.5in} (c)
%     \end{tabular}
    \caption{Monitoring statistics of the compared methods for the LDED monitoring application, with the true change-point $t^*=30$ dotted in red. (Left) and (Middle) show the run length posterior for BLAST and un-BLAST, with its maximum a posterior run length $r_t^{\rm MAP}$ marked by solid lines. (Right) shows the monitoring statistics from existing methods.}
    \label{laser_results}
\end{figure}

In what follows, we use the two images in Figure \ref{fig:lded} (right), both with $42 \times 46$ pixels, as the pre-change and post-change images. Noisy image data are then generated with i.i.d. $\mathcal{N}(0,1.1^2)$ noise, with the true change-point occurring at $t^* = 30$. The same methods are implemented and compared as in earlier experiments. For BLAST, the DGMRF prior is elicited from a separate set of $N = 50$ noisy images sampled from the pre-change image. 

Figures \ref{laser_results} (left) and (middle) show the run length posteriors for BLAST and un-BLAST. We see again that the BLAST run length posterior increases steadily with high probability prior to the change-point, then dips down to zero with high probability after the change-point, as desired. For un-BLAST, however, its run length posterior is highly uncertain prior to the change-point at $t^*=30$. Up until $t=15$, its run length posterior is highly diffuse with its MAP constant at zero (which is undesirable as no change has occurred yet); after this, its run length MAP steadily increases but its posterior remains highly uncertain. In retrospect, this is not too surprising, as un-BLAST does not leverage elicited image structure as prior information. After the change-point, both the run length posteriors for BLAST and un-BLAST dip quickly to zero, as desired.

Figure \ref{laser_results} (right) shows the monitoring statistics for existing methods. As before, the Hotelling-$T^2$ experiences delayed detection performance as it largely ignores the underlying image structure. The MMD performs quite well here: its monitoring statistics remain stable pre-change, and spikes up promptly after the change-point. One plausible reason is that the underlying image features (see Figure \ref{fig:lded} right) are largely circular, which is well-captured by the squared-exponential kernel. MONAD also performs quite well here. Such existing methods, however, are not Bayesian in nature, and thus do not provide the desired probabilistic monitoring uncertainty for confident change detection.

\section{Conclusion}
\label{sec:conc}

We proposed a new Bayesian onLine Structure-Aware change deTection (BLAST) method, which tackles the challenges of (i) scalable monitoring of images, (ii) capturing image structure (e.g., edges or shapes) for quick change detection, and (iii) a reliable probabilistic quantification of monitoring uncertainty. Such challenges are ubiquitous in broad applications, from surveillance to manufacturing. To address this, BLAST leverages the deep Gaussian Markov random field prior in \cite{DGMRF}, which can be elicited to capture expected image structure. With this elicited prior, BLAST then employs a carefully-constructed online Bayesian change-point procedure for scalable image monitoring. A key advantage of BLAST is that, at each time-step, the computation of its posterior run length distribution can be performed in $\mathcal{O}(p^2)$ work (where $p$ is the number of image pixels), which facilitates efficient monitoring of large images. We demonstrate the effectiveness of BLAST in a suite of numerical experiments and in two applications on street scene monitoring and metal additive manufacturing monitoring.

Despite promising developments, there are several important directions for future investigation. These directions target the broader use of BLAST for modern but perhaps less explored applications of change-point detection. One such area is in the timely \textit{novelty detection} of anomalous activity in particle physics experiments \citep{Kasieczka_2021_LHC}, which contribute to discoveries of new particles (e.g., the Higgs Boson \citep{bass2021higgs}). The key challenge there is the sheer volume of image measurement data collected, which can easily exceed terabytes of memory per second. We are exploring the incorporation of recent large-scale kernel learning methods \citep{li2023prospar,li2023trigonometric} to further improve scalability for this massive data setting; tackling this can catalyze promising high-energy physics discoveries \citep{everett2022role}. We are also investigating the extension of BLAST for non-Gaussian image measurements. Such a situation arises in high-energy physics, where experimental observables are typically measured as counts \citep{li2023additive,ehlers2022bayesian}. This non-Gaussian extension may require the integration of convex programming for real-time scalability, following recent work in \citep{ERPCA}.

\if0\blind{
% \noindent \textbf{Acknowledgements}: The authors gratefully acknowledge funding from NSF CSSI Frameworks grant 2004571 (KL, SM), NSF DMS 2210729 (KL, SM) and U.S. Department of Energy Grant no. DE-FG02-05ER41367 (SAB, JFP).
}
\fi

% \FloatBarrier
\spacingset{1.0}
\bibliography{reference}

%\bibliography{TVR_references}

\end{document}